\documentclass{jfm}
\usepackage{amscd} 

\usepackage{extarrows}
\usepackage{psfrag}
\usepackage{epstopdf}
\usepackage{color}
\usepackage{threeparttable}

\usepackage{multirow}

\usepackage{subfigure}

\usepackage{todonotes}

\usepackage{mathrsfs}
\usepackage{graphicx}
\usepackage{natbib}
\usepackage{amsmath,xfrac}
\usepackage[toc,page]{appendix}
\usepackage{mathtools}

\ifCUPmtlplainloaded \else
  \checkfont{eurm10}
  \iffontfound
    \IfFileExists{upmath.sty}
      {\typeout{^^JFound AMS Euler Roman fonts on the system,
                   using the 'upmath' package.^^J}%
       \usepackage{upmath}}
      {\typeout{^^JFound AMS Euler Roman fonts on the system, but you
                   dont seem to have the}%
       \typeout{'upmath' package installed. JFM.cls can take advantage
                 of these fonts,^^Jif you use 'upmath' package.^^J}%
      }
  \else
  \fi
\fi

\ifCUPmtlplainloaded \else
  \checkfont{msam10}
  \iffontfound
    \IfFileExists{amssymb.sty}
      {\typeout{^^JFound AMS Symbol fonts on the system, using the
                'amssymb' package.^^J}%
       \usepackage{amssymb}%
       \let\le=\leqslant  
         
      }{}
  \fi
\fi

\ifCUPmtlplainloaded \else
  \IfFileExists{amsbsy.sty}
    {\typeout{^^JFound the 'amsbsy' package on the system, using it.^^J}%
     \usepackage{amsbsy}}
    {}
\fi

\newcommand{\EQ}{\begin{equation}}
\newcommand{\EN}{\end{equation}}
\newcommand{\EQA}{\begin{eqnarray}}
\newcommand{\ENA}{\end{eqnarray}}
\newcommand{\EQL}{\begin{align}}
\newcommand{\ENL}{\end{align}}

\newsavebox{\astrutbox}
\sbox{\astrutbox}{\rule[-5pt]{0pt}{20pt}}

\newcommand{\be}{\begin{equation}}
\newcommand{\ee}{\end{equation}}

\title[LES of Taylor-Couette flow]
{ Large-eddy simulation and modeling of  Taylor-Couette flow with an outer stationary cylinder}

\author[ W. Cheng, D.I.  Pullin and R. Samtaney ]%
{W.\ns C\ls H\ls E\ls N\ls G$^{1,2}$%
  \ns
, D.\ns I.\ns P\ls U\ls L\ls L\ls I\ls N$^2$%
  \ns
,  R.\ns S\ls A\ls M\ls T\ls A\ls N\ls E\ls Y$^1$%
  \break
 }
\affiliation{$^1$Mechanical Engineering, Physical Science and Engineering Division,
King Abdullah University of Science and Technology,
Thuwal, Saudi Arabia, 23955-6900  \\[\affilskip]
$^2$Graduate Aerospace Laboratories, California Institute of Technology, CA, 91125, USA
}

\pubyear{2008}
\volume{xxx}
\pagerange{119--126}
\date{?? and in revised form ??}

\begin{document}

\large

\maketitle

\begin{abstract}

We present wall-resolved  large-eddy simulations (LES) of the incompressible Navier-Stokes equations together with  empirical modeling  for {turbulent} Taylor-Couette {(TC)}  flow where the inner cylinder is rotating with angular velocity $\Omega_i$ and the outer cylinder is stationary.  With $R_i, R_o$ the inner and outer radii respectively, the radius ratio is $\eta = 0.909$. The subgrid-scale (SGS) stresses are represented using the stretched-vortex subgrid-scale model  while the flow is resolved close to the wall.  LES is implemented in the range $ Re_i = 10^5 - 3\times10^6$ where $Re_i  = \Omega_i\,R_i\,d/\nu$ and $d= R_o-R_i$ is the cylinder gap. It is shown that the LES can
capture the salient features of the TC flow, including the quantitative behavior of span-wise Taylor rolls, the log-variation in the mean velocity profile and the angular momentum redistribution due to the presence of Taylor rolls.  A simple empirical model of the turbulent, TC flow is developed consisting of near-wall, log-like turbulent wall layers separated by an annulus of constant angular momentum. The model is closed by a proposed  scaling relation concerning the thickness of the wall layer on the inner cylinder. Model results include  the Nusselt number $Nu$ (torque required to maintain the flow) and various measures of the wall-layer thickness as a function of both the Taylor {number}  $Ta$ and $\eta$.  These agree reasonably with experimental measurements, direct numerical simulation (DNS) and the present LES over a  range of both $Ta$ and $\eta$. In particular, the model shows that, at fixed $\eta<1$, $Nu$ grows like $Ta^{1/2}$ divided by the square of the Lambert, (or Product-Log)  function of a variable proportional to $Ta^{1/4}$. This cannot be represented by a power law dependence on $Ta$. At the same time the wall-layer thicknesses reduce slowly in relation to the cylinder gap. This suggests an asymptotic, very large  $Ta$ state consisting of constant angular momentum in the cylinder gap with  $u_\theta = 0.5\,\Omega_i\,R_i^2/r$, where $r$ is the radius,  with vanishingly thin turbulent wall layers at the cylinder surfaces.  An extension of the model to rough-wall turbulent wall  flow at the inner cylinder surface is described. This shows an asymptotic, fully rough-wall state where the torque is  independent of $Re_i/Ta$, and  where $Nu\sim Ta^{1/2}$.
\end{abstract}

\def\diam{$\diamond$\nobreak\ }
\def\square{${\vcenter{\hrule height .4pt
        \hbox{\vrule width .4pt height 3pt \kern 3pt
        \vrule width .4pt}
        \hrule height .4pt}}$\nobreak\ }
\def\circle{$\circ$\nobreak\ }
\def\trian{\raise 1.25pt\hbox{$\scriptscriptstyle\triangle$}\nobreak\ }
\def\tinynabla{\raise 1.25pt\hbox{$\scriptscriptstyle\nabla\textstyle$}\nobreak\ }
\def\drawline#1#2{\raise 2.5pt\vbox{\hrule width #1pt height #2pt}}
\def\spacce#1{\hskip #1pt}
\def\solid{\drawline{24}{.5}\nobreak\ }
\def\bdash{\hbox{\drawline{4}{.5}\spacce{2}}}
\def\dashed{\bdash\bdash\bdash\bdash\nobreak\ }
\def\lbdash{\hbox{\drawline{7}{.5}\spacce{2}}}
\def\longdashed{\lbdash\lbdash\lbdash\lbdash\nobreak\ }
\def\chndot{\hbox {\drawline{9.5}{.5}\spacce{2}\drawline{1}{.5}\spacce{2}\drawline{9.5}{.5}}\nobreak\ }

\def\dotdot{\hbox {\drawline{1.0}{1.0}\spacce{2.0}\drawline{1.0}{1.0}
\drawline{1.0}{1.0}\spacce{2.0}\drawline{1.0}{1.0}
\drawline{1.0}{1.0}\spacce{2.0}\drawline{1.0}{1.0} }\nobreak\ }

\def\chndotdot{\hbox {\drawline{9.5}{.5}\spacce{2}\drawline{1}{.5}\spacce{2}\drawline{1}{.5}\spacce{2}\drawline{9.5}{.5}}\nobreak\ }

\newcommand*\sq{\mathbin{\vcenter{\hbox{\rule{1.2ex}{1.2ex}}}}}
\section{Introduction}
\smallskip

Taylor-Couette (TC) flow of a viscous fluid in the annular gap between two concentric cylinders,
where one or  both cylinders are  rotating, is a classical turbulent flow that exhibits
 interesting shear-flow phenomena ~\cite[]{taylor1923stability,grossmann2016high}.  TC flow is perhaps  more experimentally  accessible than the related plane-Couette (PC) flow (e.g.  ~\cite{Pirozzoli2014Turbulence}) owing to the cylindrical geometry and the convenience of torque measurement.  The parameter space covered by most experimental and computation studies of TC flow includes independent Reynolds numbers associated with inner and/or  outer rotation angular velocities,  respectively, a radius ratio   and the distinction between co-rotating and counter-rotating cylinder motion.
 {TC flow can be defined by  two independent length  and two independent velocity scales. For incompressible TC flow of a Newtonian fluid, this gives three dimensionless numbers  which typically are the radius ratio $\eta=R_i/R_o$ and the  inner and outer Reynolds numbers $Re_i = \Omega_i\,R_i\,d/\nu$ and
 $Re_o = \Omega_o\,R_o\,d/\nu$ respectively. Here $d= R_o-R_i$ with $R_i$ the radius of the inner cylinder and $R_o$ that of the outer cylinder,
$\Omega_i$, $\Omega_o$ are rotation angular velocities of the inner and  outer cylinders respectively and $\nu$  the kinematic viscosity of the fluid.
Alternative specifications are sometimes useful  such as  the Taylor number $Ta$ for the purpose of  analogy with Rayleigh-B{\'e}nard flow, and  $Re_w$, a Reynolds number based on the standard deviation of the radial velocity \cite[]{Huisman2012}.
}

Figure 1  of ~\cite{andereck1986flow}, reproduced as figure 2 of  ~\cite{grossmann2016high}, shows a classification of observed flow types in a $Re_i-Re_o$ plane up to moderate Reynolds number $Re_i\approx 2000$. These include wavy vortex flow, modulated waves, spiral turbulence and Taylor vortex flow among others.
 The review article of ~\cite{grossmann2016high} surveys and summarizes research on these  fluid-dynamical phenomena associated with TC flow. At large sufficiently large $Re_i$ (or  large Taylor number $Ta$, which for $Re_o=0$  is proportional to $Re_i^2$) they point out that the near-wall layers on the cylinder walls become turbulent signalling a transition toward an ``ultimate regime''. Here the two cylinder  wall layers  appear to conform to the classical law of the wall and are separated by a region of bulk flow that is dominated by large-scale unsteady phenomena such as span-wise Taylor roll structures. Experiments { at different $\eta$ } have been conducted in this regime in the range  $Ta = 10^{11}-10^{13} $~\cite[]{merbold2013torque,van2011torque,van2012optimal}. These show a Nusselt number $Nu$ - the ratio of torque required to maintain the motion to the laminar-flow torque -  variation with $Ta$  that can be reasonably approximated over this range by $Nu \sim Ta^p$ where $p$ is less than $0.5$.

   \cite{ostilla2016near} report  direct numerical simulation (DNS)   of TC flow with the outer cylinder stationary
  up to $Re_i = 3\times 10^5$ for $\eta = 0.909$. This corresponds to  a maximum Taylor number $Ta = 9.969\times 10^{10} $. Their results show two Taylor rolls  and also demonstrates  that the bulk region separating the cylinder wall layers consists of a region of almost constant mean angular momentum density that corresponds to the average of that corresponding to the two cylinder radii and angular velocities.

 Presently we investigate Taylor-Couette flow at relatively large Reynolds numbers using the numerical technique of large eddy simulation (LES).  Our aim in part is to provide data at larger $Re_i$ than is presently  available from DNS as a prelude to wall-modeled LES at even larger $Re_i$.  We utilize $\eta = 0.909$ with $Re_o=0$ and  $Re_i= 10^5,3\times 10^5,6\times 10^5, 10^6, 3\times 10^6$, with a maximum Taylor number $Ta = 9.969\times10^{12}$.
 In \S \ref{LESdescription} we outline the numerical method and the subgrid-scale model for our wall-resolved LES. This is followed in \S  \ref{LESResults} by an account of the present LES results.  Good agreement with the DNS of  \cite{ostilla2016near} is obtained at our lower $Re_i$ . The higher $Re_i$ results show a clear log-like profile for the wall layer of the inner cylinder. All LES reproduce a bulk inner region with almost constant azimuthal-spanwise averaged angular momentum. \S \ref{FlowModel}  describes   an empirical, one dimensional (radial) model for the mean TC flow for relatively large $Re_i$ and $Re_o=0$.  This comprises wall bounded regions with law-of-the wall mean velocity profiles together with a uniform angular momentum central region.  The model is closed with a scaling hypothesis concerning the relative thickness of the inner wall layer.  It is shown that the model gives satisfactory agreement  with experiment, DNS and LES for several important mean-flow parameters. In \S \ref{RoughWalls}
an extension of the model to rough-wall  layers is described, while concluding remarks are presented in \S \ref{Conclusion}.

\section{Large-eddy simulation}
\label{LESdescription}

\smallskip
\subsection{Numerical method}
The governing equations for LES of incompressible viscous flow are derived by formally applying a spatial filter onto the Navier-Stokes equation. In Cartesian co-ordinates  $x_i$, $i=1,2,3$ these are
\begin{equation}
\dfrac{\partial \tilde{u}_i}{\partial t} + \dfrac{\partial \tilde{u}_i\, \tilde{u}_j}{\partial x_j}
= -\dfrac{\partial \tilde{p}}{\partial x_j} + \nu \frac{\partial^2 \tilde{u}_i}{\partial x_j^2} - \dfrac{\partial T_{ij}}{\partial x_j}, \quad \dfrac{\partial \tilde{u}_i}{\partial x_i} = 0.\label{LES-Eqns}
\end{equation}
with $\tilde{u}_i$ the  filtered velocity and
$\tilde{p}$  the filtered pressure, and $T_{ij}= \widetilde{u_iu_j}- \widetilde{u_i}\widetilde{u_j}$ denotes the effect of subfilter scales on the resolved-scale motion. In practice,  this is represented on a computational grid using a  subgrid-scale (SGS) model.
For convenience  we will also utilize  $(x,y,z)$ as  Cartesian coordinates with $(\tilde{u},\tilde{v},\tilde{w})$ as the corresponding filtered velocity components.
Additional coordinate systems  are also used. Cylindrical coordinates ($\theta,y,r$) with velocity components ($u_\theta,u_y, u_r$) are convenient for  diagnosing results. General  curvilinear coordinates $(\xi, y, \eta)$ will be described for  the implementation of the numerical method.

\subsection{Numerical method}

In the  curvilinear coordinate system, the (formally) filtered governing equations in conservation-law form can be written as ~\cite[]{Zang1994}
\begin{equation}
\frac{\partial U^m}{\partial \xi^m} = 0 , \ \ \
\frac{ J^{-1} \partial \tilde{u_i}}{\partial t} +\frac{\partial F_i^m}{ \partial \xi^m } = 0,
\end{equation}
where $U^m$ and $F^m$ are given as
\begin{equation}
U^m = J^{-1} \frac{ \partial \xi^m }{ \partial x_i} \widetilde{u_i} \  \ \text{and} \  \  \
F_i^m   = U^m \widetilde{u_i}  + J^{-1} \frac{\partial \xi^m}{\partial x_j} T_{ij}
 + J^{-1} \frac{\partial \xi^m}{\partial x_i} \tilde{p}
- \nu G^{mn} \frac{\partial \widetilde{u_i} }{\partial \xi^n},
\end{equation}
respectively where  $J^{-1}$ is the inverse of the Jacobian and $G^{mn}$ is the mesh skewness tensor defined as:
\begin{equation}
J^{-1} = \det \left( \frac{\partial x_i}{\partial \xi^j} \right),  \ \ \
G^{mn} = J^{-1} \frac{\partial \xi^m}{\partial x_j} \frac{\partial \xi^n}{\partial x_j}.
\end{equation}

A semi-implicit fractional step method was used to solve the governing equations with successive solution of modified Helmholtz equations that results from  implicit treatment of the viscous terms, pressure Poisson equation and the velocity correction step. Integration  in time is implemented using an Adams-Bashforth method for  explicit terms and  Crank-Nicolson  for  implicit terms.  A parallel multi-grid solver with a line-relaxed Gauss-Seidel iteration method is used for numerical solution of the Poisson equations.  The spatial discretization of the nonlinear term utilizes
a fourth-order energy-conservative scheme of the skew-symmetric form by~\cite{Morinishi1998}, while for all other terms are discretized using a fourth-order central difference scheme.
 The present code framework has been verified and validated for several flows that
 include  flow over an airfoil using both DNS~\cite[]{Zhang2015} and  wall-modeled LES \cite[]{Gao2019} and
 wall-resolved LES of flow over a circular cylinder in  different configurations ~\cite[]{cheng2017large,Cheng2018,Cheng2018b} .
 All LES described presently  were  performed on the Cray XC40 supercomputer Shaheen at KAUST.

\subsection{Stretched vortex SGS model}

We utilize the  stretched-vortex (SV) SGS model \cite[]{Misra1997,Voelkl2000,Chung2009} in regions away from the wall. This is a structure-based model
where the subgrid flow is represented  by tube-like, spiral vortices \cite[]{Lundgren1982} stretched by the rate-of-strain tensor of  the local resolved-scale flow.  Inside a computational cell there exists an (virtual) SGS  vortex with direction vector $\textbf{\textit{e}}^v$  resulting in the subgrid stress
\begin{equation}
T_{ij} =
(\delta_{ij} - \textbf{\textit{e}}_i^v\textbf{\textit{e}}_j^v) K,
\end{equation}
where $K$ is the subgrid kinetic energy, expressed as  integral of the SGS energy spectrum \cite{Lundgren1982}as
\begin{equation}
 K=\int_{k_c}^\infty E(k) dk
= \frac{\mathscr{K}'_0}{2} \Gamma\left[-1/3,  \frac{2\nu k_c^2}{3|\tilde{a}|}\right],
\end{equation}
where  $\Gamma[..,..]$ is the incomplete gamma function,  $k_c=\pi/\Delta_c$ is the cutoff wavenumber,
$\tilde{a}=\textbf{\textit{e}}_i^v \textbf{\textit{e}}_j^v \widetilde{S}_{ij}$
is the resolved-scale stretching along the subgrid vortex
with $\widetilde{S}_{ij}$ the resolved-scale, rate-of-strain tensor. The $\textbf{\textit{e}}_j^v$ are
aligned with the principal extensional eigenvector of  $\widetilde{S}_{ij}$ while
the parameter $\mathscr{K}'_0$ can be calculated  dynamically from
the resolved-scale velocity using a matching procedure as $\mathscr{K}'_0= \langle F_2
\rangle/ \langle Q(\kappa_c,d)\rangle$ where $\langle  \rangle$ denotes an averaging  strategy,
 computed as the arithmetic mean of $26$ neighboring points { and $\kappa_c = k_c\,(2\,\nu/3|\tilde{a}|)^{1/2}$ ~\cite[]{Chung2009}}.  The second-order local structure function  of the resolved-scale velocity field is
 $F_2$  and $Q(\kappa_c,d)$ is calculated using an asymptotic approximation with $d=r/\Delta_c$ where
$r$ the distance from neighbor point to the vortex axis. The SV SGS model is implemented in a strictly local setting and does not require either local isotropy of homogeneity in one or more co-ordinate directions.  For details see \cite{Misra1997,Voelkl2000, Chung2009}.
 The present LES is ``wall-resolved'' meaning that the wall-normal grid size at the wall is of order the local viscous wall scale $u_\tau/\nu$ where $u_\tau \equiv \sqrt{|\tau_w|/\rho}$ is the friction velocity with $|\tau_w|$ the magnitude of the wall shear stress and $\rho$ the constant fluid density.

\begin{figure}
\begin{center}	
\includegraphics[height=7.5cm]{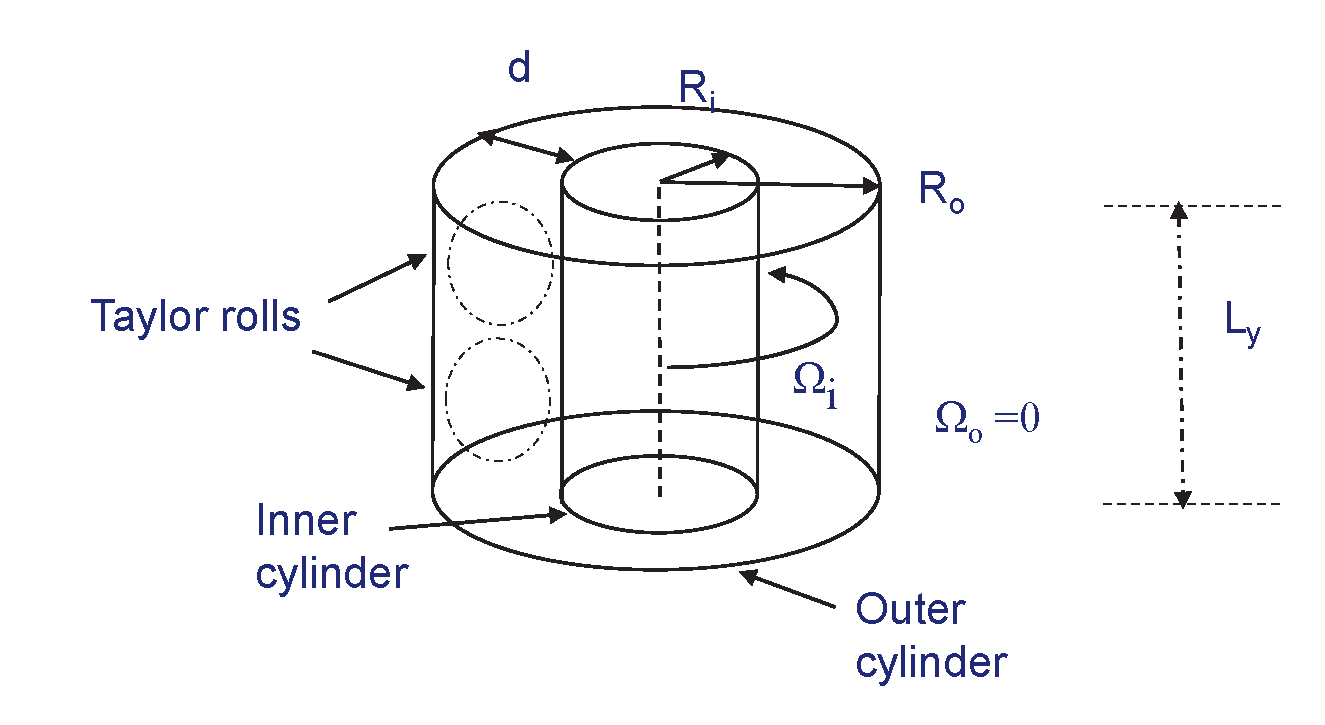}
\caption{{Flow configuration for Taylor-Couette flow with rotating inner cylinder { ($\Omega_i\ne 0$)}
and stationary outer cylinder {($\Omega_o=0$)}.
{ $R_i$ is the radius of the inner cylinder, $R_o$ is the radius of the outer cylinder, $d=R_o-R_i$.  } }}
\label{TCDiagramA}
\end{center}
\end{figure}


\begin{table}
\begin{center}
\begin{tabular}{p{1.2cm}p{1.0cm}p{1.0cm}p{1.0cm}p{2.0cm}p{2.0cm}p{1.2cm}p{1.2cm}p{1.2cm}}
\hline
 $Re_i$   & $N_{\theta}$ & $N_r$ & $N_y$ &  $Ta$&$Re_{\tau_i}$ &  $r_i \Delta \theta^+ $ & $\Delta r_{min}^+$ & $\Delta y^+$   \\  \hline
 $1\times 10^5$   & 256 & 256 & 1024 &$1.108\times 10^{10}$& $1.400\times 10^3$ &  34.3  &  0.75  & 5.74   \\ 
 $3\times 10^5$   & 512 & 512 & 1536 &$9.969\times 10^{10}$& $3.908\times10^3$ &47.9 &  0.54 &   10.7  \\ 
 $6\times 10^5$   & 1024 & 512 & 2048 &$3.988\times 10^{11}$& $7.289\times 10^3$ & 44.7 &  0.51 &   14.9  \\
 $1\times 10^6$   & 2048 & 1024 & 4096  & $1.108\times 10^{12}$&$1.125\times 10^4 $ & 68.9 &  0.77  &  11.5  \\  
 $3\times 10^6$   & 2048 & 1024 & 4096  & $9.969\times 10^{12}$&$3.178\times 10^4 $ & 97.4  & 1.07 &   32.6  \\  \hline
\end{tabular}
\end{center}
\caption\small{{Parameters for LES  at varying $Re_i$ with $\eta = 0.909$.
{ $N_\theta$, $N_r$ and $N_y$ are mesh numbers employed in the azimuthal direction, radial direction and spanwise direction, respectively. }
For all cases,  the domain size  is a sector of $\Delta \theta =\pi/10$ in the azimuthal $\theta$ direction and is $L_y = 2\pi d/3$ in the span-wise $y$ direction.
{$Re_{\tau_i} = u_{\tau i}d/2\nu$ as calculated from the LES. Mesh sizes in   viscous wall scaling are $r_i\Delta \theta^+$ and $\Delta y^+$ respectively, uniform  in azimuthal  and spanwise directions. $\Delta r^+_{min}$ is the minimal near wall mesh size. }   }}
\label{table:cases}
\end{table}

\begin{figure}
\begin{center}	
\includegraphics[height=6.45cm]{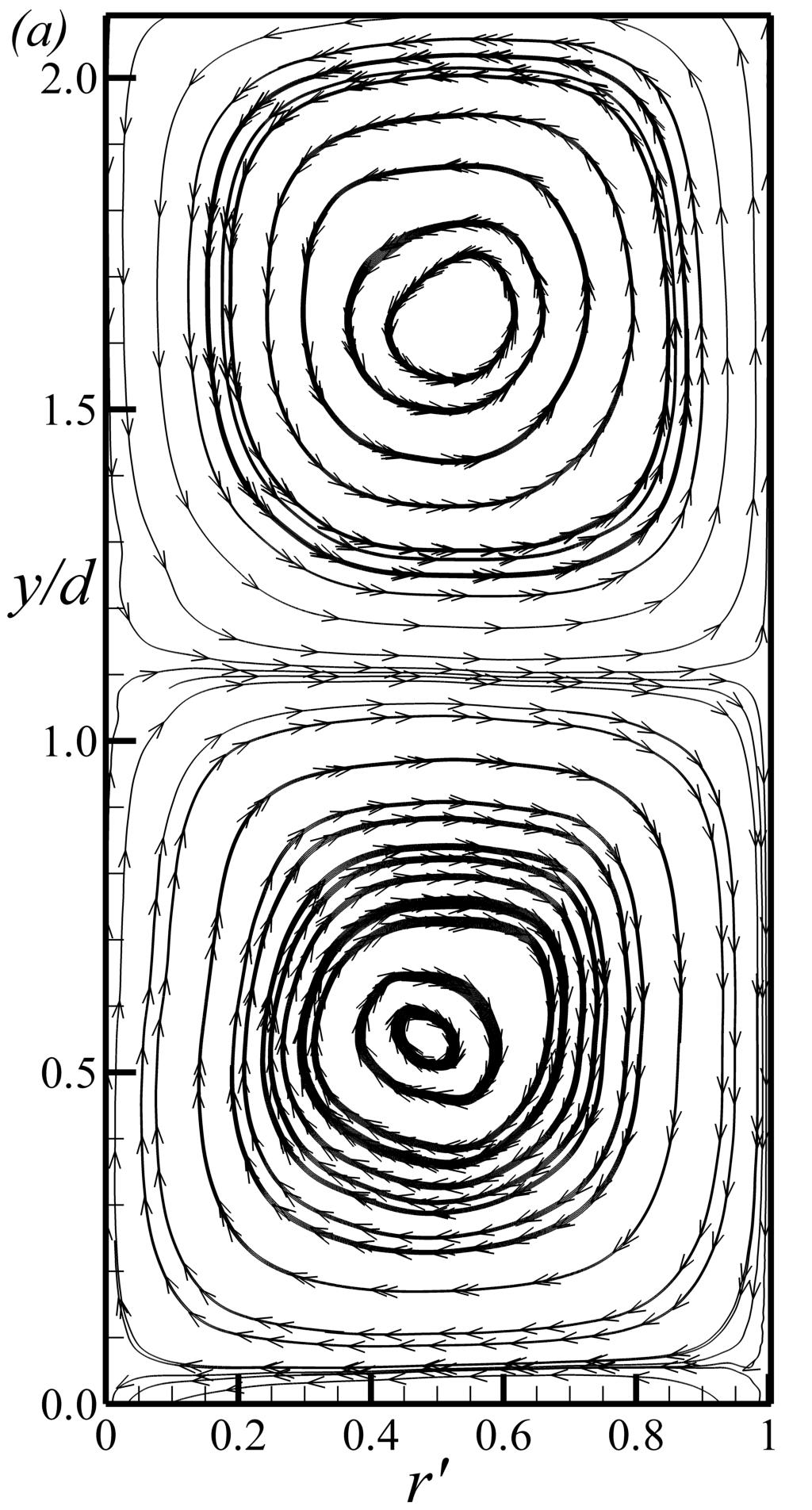}
\includegraphics[height=6.5cm]{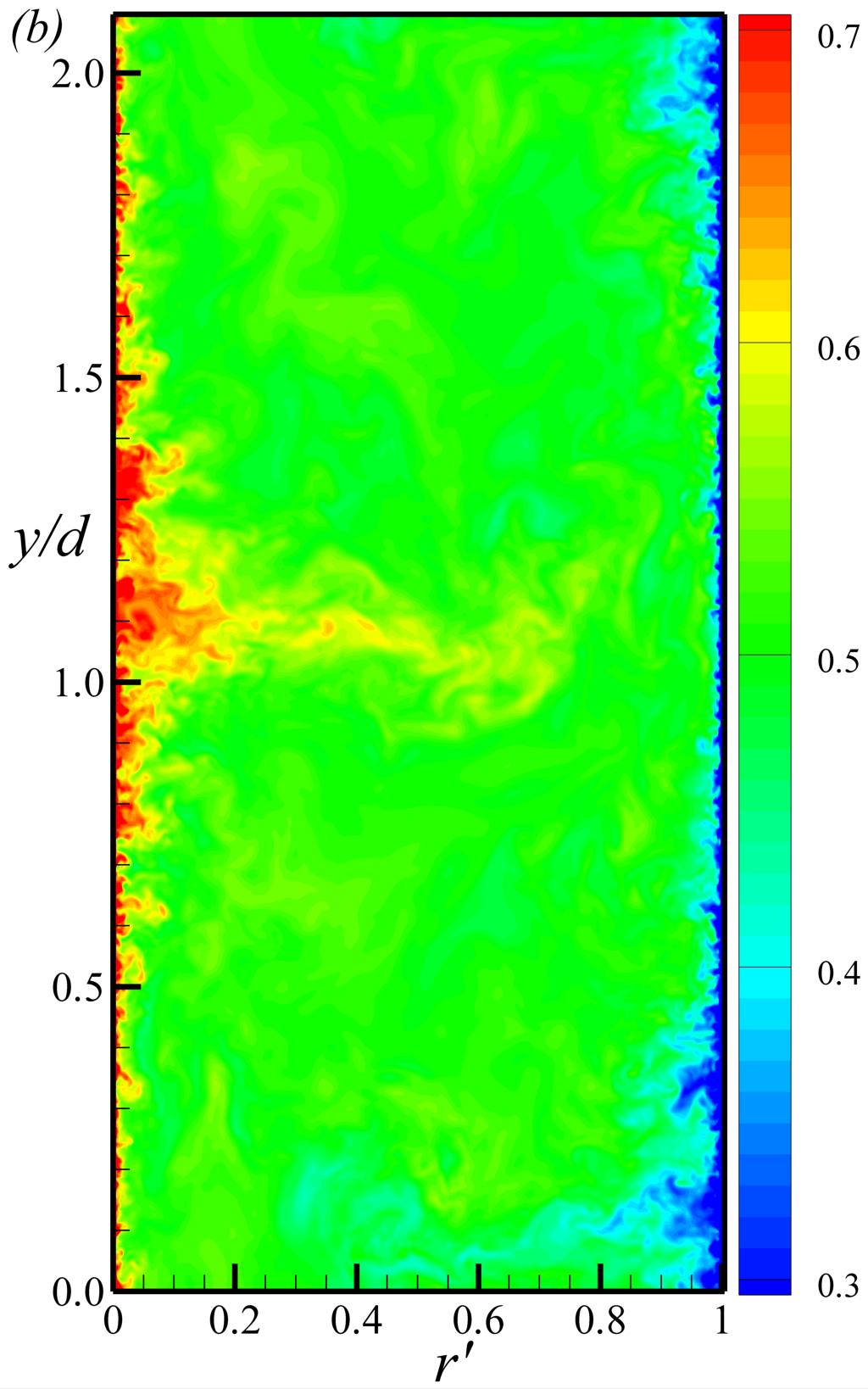}
\includegraphics[height=6.5cm]{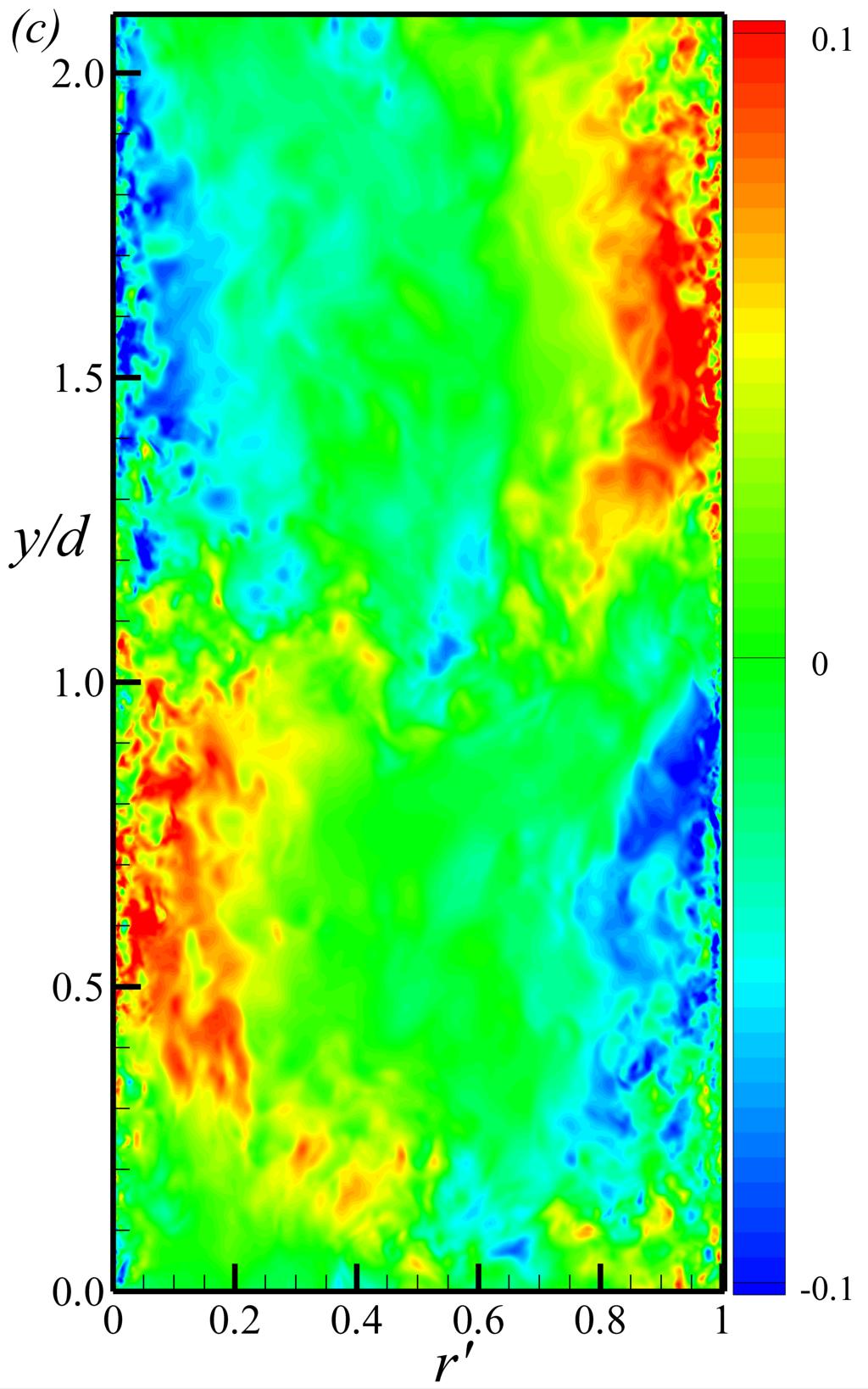}
\caption{Visualization of an instantaneous flow field in a radial-spanwise plane: $r' = (r-R_i)/d, \theta$. $Re_i=10^5$, $\eta = 0.909$.
(a), streamlines of the azimuthal-averaged flow field $(u_r, u_y)$;
(b), instantaneous azimuthal velocity field at mid-span plane;
(c), instantaneous span-wise velocity field at mid-span plane.}
\label{figure1}
\end{center}
\end{figure}

\section{LES results}
\label{LESResults}

\subsection{Cases implemented}
\smallskip
In the present LES of Taylor-Couette flow, the outer cylinder is stationary; $\Omega_o = 0$,
as shown in the flow configuration of  figure \ref{TCDiagramA}.
The relevant  dimensionless parameters defining the flow are  then the radius ratio $\eta$ and the inner-cylinder Reynolds number $Re_i$.
The  inner-cylinder friction Reynolds number  is
\begin{equation}
 Re_{\tau_i} = \dfrac{u_{\tau_i}\,d}{2\,\nu},
\label{Parameters}
\end{equation}
 where $u_{\tau_i} = \sqrt{\tau_{i,w}/\rho}$ is the inner cylinder friction velocity, with  $\tau_{i,w}$ the shear stress at the wall.

 Generally, and in experiments, both $\eta, Re_i$ are fixed but $Re_{\tau_i}$ must be determined by measurement, numerical simulation, theory or modeling.  The torque that must be applied to the outer cylinder to sustain the motion is $T=2\pi R_i L_y R_i\tau_{i,w}$.  In the present LES we use $\eta = 0.909$ and vary  only $Re_i$.
In cylindrical $(r,\theta,y)$ co-ordinates, the computational domain is a sector of angle $\pi/10$ in the $\theta$-direction, which is a well-accepted domain size in DNS simulation by \cite{Ostilla2015,ostilla2016near}. In the span-wise direction the domain length is $L_y = 2\pi d/3$.
  Periodic boundary conditions are implemented in both $\theta$ and $y$. Grid spacing  is uniform  both   $\theta$ and span-wise $y$ but is stretched in the $r$ direction.

The present study focuses on the flow  behavior at relatively high $Re_i$.
For numerical verification we utilize DNS  $Re_i$ at $10^5$ and $3\times 10^5$  \cite[]{ostilla2016near}.
LES at higher $Re_i$  {up to  $3\times 10^6$} are also presented.  Parameters for the five LES performed are
listed in table \ref{table:cases}.  These include the number of grid cells in each direction $(N_\theta, N_r, N_y)$ and the Taylor number, which is defined as  ~\cite[]{grossmann2016high}
\begin{equation}
Ta = \frac{(1+\eta)^4}{64\,\eta^2}\,\frac{(R_o-R_i)^2(R_o+R_i)^2\,(\Omega_i-\Omega_o)^2}{\nu^2}.
\label{TaNum}
\end{equation}
For $\Omega_o=0$  this becomes
\begin{equation}
Ta =  \dfrac{(1+\eta)^6 }{64\,\eta^4}\,Re_i^{2} .
\label{TaNumII}
\end{equation}
With $\eta = 0.909$ this is $Ta = 1.1076\, Re_i^2$.

For   the purpose of defining  averaged quantities the flow is assumed to be statistically stationary in time over a sufficiently long time period following initial transients, and spatially homogeneous in the $\theta$ direction only.  In the span-wise direction the   flow is  generally non-homogenous  owing the presence of Taylor rolls.
Starting from a scalar field  $\phi(\theta,y,r,t)$,  `` $\hat{..}$'' denotes an average of a space-time dependent quantity  in both time and the azimuthal ($\theta$) direction,   resulting in $\hat{\phi}(y,r)$, while  ``$\bar{..}$ " denotes an additional span-wise average  of  $\hat{\phi}(y,r)$, resulting in $\overline{\hat{\phi}}(r)$ .

 We denote the mean radial velocity {in the laboratory frame of reference}  as ${\textbf{U}}(r) = \overline{ \hat{\textbf{u}  }}$.
For computation of turbulent statistics, a velocity  fluctuation is first defined as $ \textbf{u}'(\theta,y,r,t)= \textbf{u}(\theta,y,r,t)- \hat{\textbf{u}}(y,r)$.
Then the  turbulent intensity is computed as $ R_{\textbf{u}\textbf{u}}  = \overline{ \widehat{\textbf{u}'\textbf{u}'  }}$.
In displaying data, both mean velocity and turbulent intensities are scaled using $u_{\tau_i}$.
With focus on the inner cylinder,  following \cite{ostilla2016near},
 we use a scaled and adjusted mean azimuthal velocity  $U^+ = U(r)/ u_{\tau i} $   with $ U (r)= \Omega_i\,R_i -  {U_\theta}(r)$ and
scaled turbulent intensities $(u_j u_j)^+ =R_{u_j u_j} / u^2_{\tau i} $ with $j$ denoting $\theta$, $y$ or $r$.
These comprise the one-point   turbulent statistics in present LES study.

\subsection{Verification with DNS at $Re_i=10^5,\, 3\times 10^5$}
\smallskip

\smallskip
We document verification of our LES  using  the benchmark  DNS of \cite{ostilla2016near} at both  $Re_i=10^5$ and   $Re_i=3\times 10^5$, the  latter being  highest $Re_i$. Comparisons mainly include mean velocity profiles and also turbulent intensities.   We can observe the flow field either in the sector domain in Cartesian coordinates, or in a developed $(r,\theta)$ domain in  cylindrical coordinates.

Figure \ref{figure1}   shows diagnostics of the  flow field at $Re_i=10^5$ viewed in  an $r-y$ or radial-spanwise  plane, where
coordinates are scaled using the cylinder gap $d$.  In the $r$ direction,   a dimensionless length scale is defined based on the distance off the inner cylinder,
as $r'=(r-R_i)/d$.  $r' = 0,1$  correspond  to the inner, rotating cylinder and  the outer static cylinder respectively.
The left sub-panel in figure \ref{figure1} shows   streamlines of the  stream-wise-averaged, instantaneous flow field in an $(r-y)$ plane.  One pair of  Taylor rolls is observed.
The center and right-hand panels  show color coded images of the instantaneous azimuthal component $u_\theta$ and  the span-wise velocity component $u_y$, respectively.

\begin{figure}
\includegraphics[width=6.7cm]{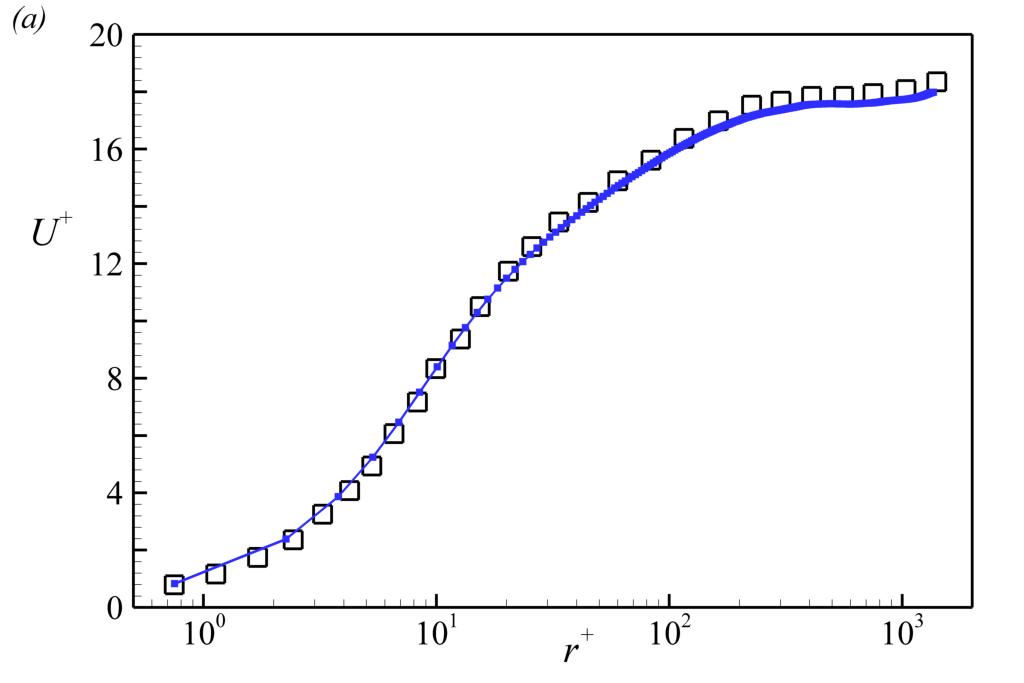}
\includegraphics[width=6.7cm]{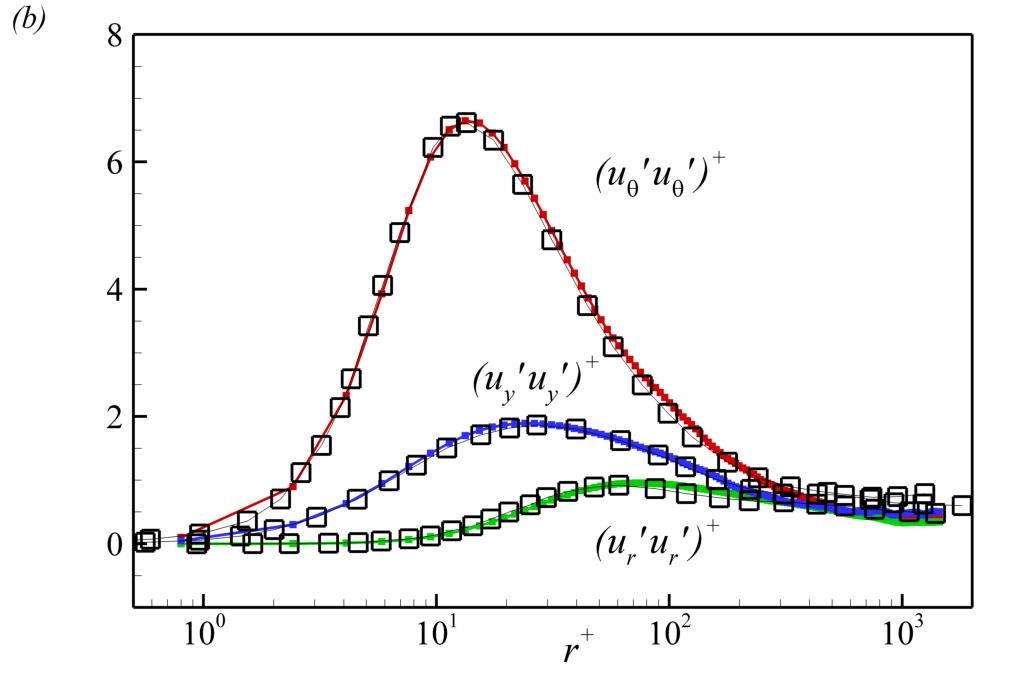}
\includegraphics[width=6.7cm]{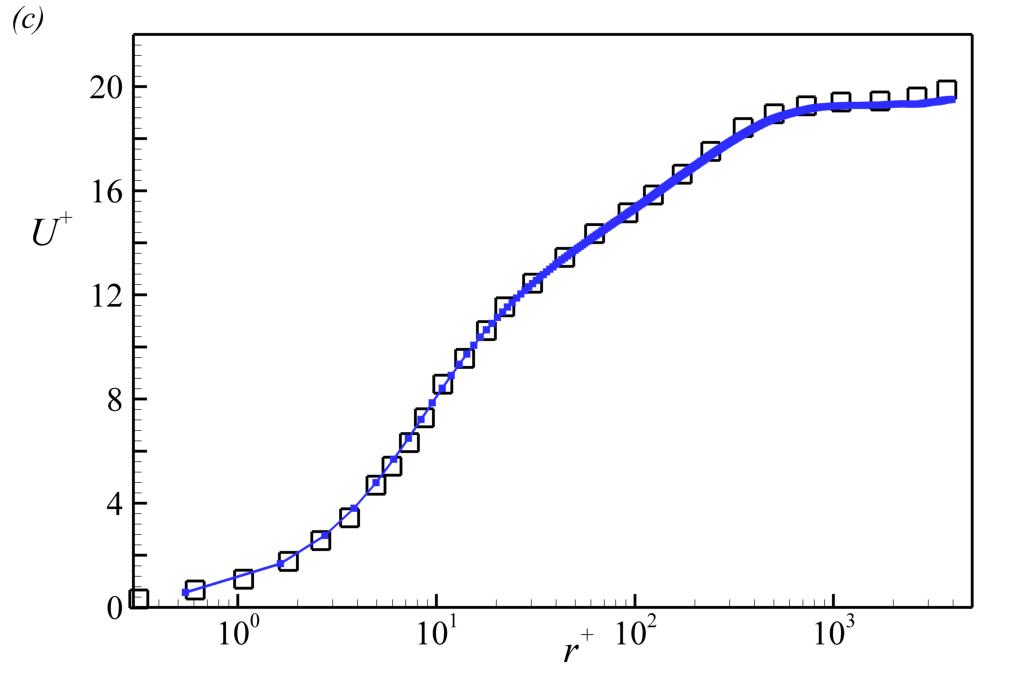}
\includegraphics[width=6.7cm]{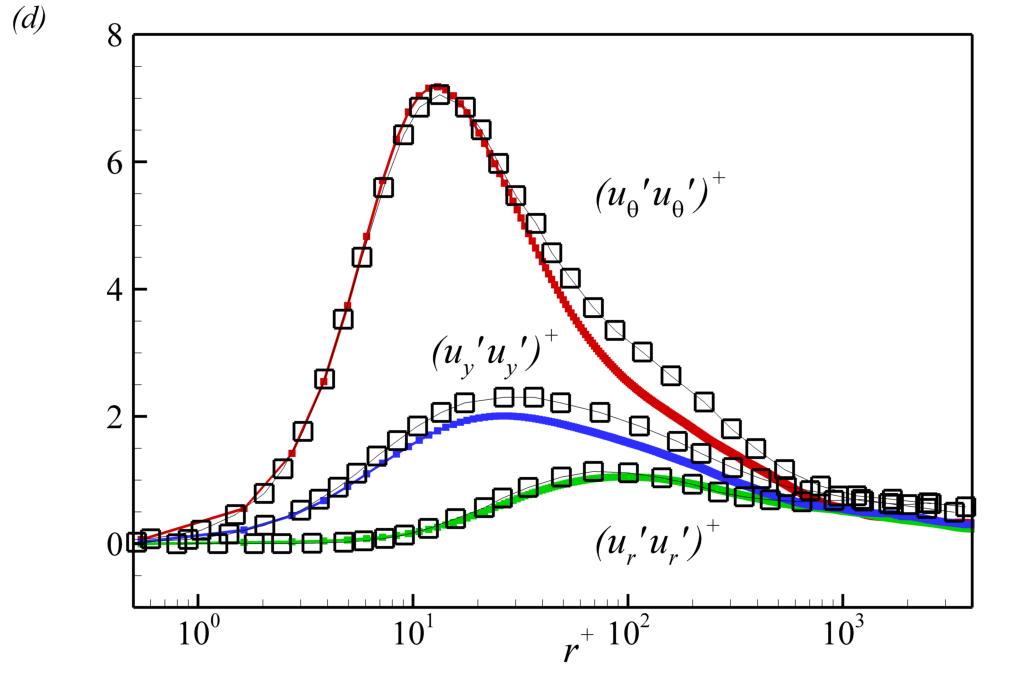}
\caption{Comparison of LES with  DNS by \cite{ostilla2016near}.
(a), mean flow velocity profiles $U^+$ at  $Re_i=10^5$; (b), turbulent intensities $(u'_\theta u'_\theta)^+, (u'_y u'_y)^+, (u'_r u'_r)^+$ at  $Re_i=10^5$.
(c),  $U^+$ at  $Re_i=3\times10^5$; (b), turbulent intensities  at  $Re_i=3\times10^5$.
Square symbols: DNS by \cite{ostilla2016near}. Solid lines with filled squares: present LES. }
\label{figure3}
\end{figure}

 In Figure  \ref{figure3},   radial profiles of the mean azimuthal velocity   and  turbulent intensities  are shown for $Re_i = 10^5,3\times 10^5$. Both mean and turbulent intensities are  scaled with  $u_{\tau_i}$,
  with $U^+$  in left panels and
   $(u'_\theta u'_\theta)^+   $,
 $(u'_y u'_y)^+  $
 and $(u'_r u'_r)^+ \ $ in the right panels
 versus the scaled length $r^+ = (r-R_i)/l^+$ with $l^+ = \nu/u_{\tau_i}$.
 The LES mean-velocity profile   shows satisfactory agreement with the direct numerical simulation (DNS) by ~\cite{ostilla2016near} for both $Re_i$. A clear log variation is evident in  both $U^+$ versus $r^+$ plots.
The present LES mesh is substantially coarser than  required for DNS.  For $Re_i=10^5$, the total LES mesh count  $N=N_\theta N_r N_z$ is $1/32$ of that for the corresponding DNS, while for $Re_i=3\times10^5$, this fraction is $1/24$.

\subsection{ Mean profiles}
\smallskip

\begin{figure}
\includegraphics[width=6.7cm]{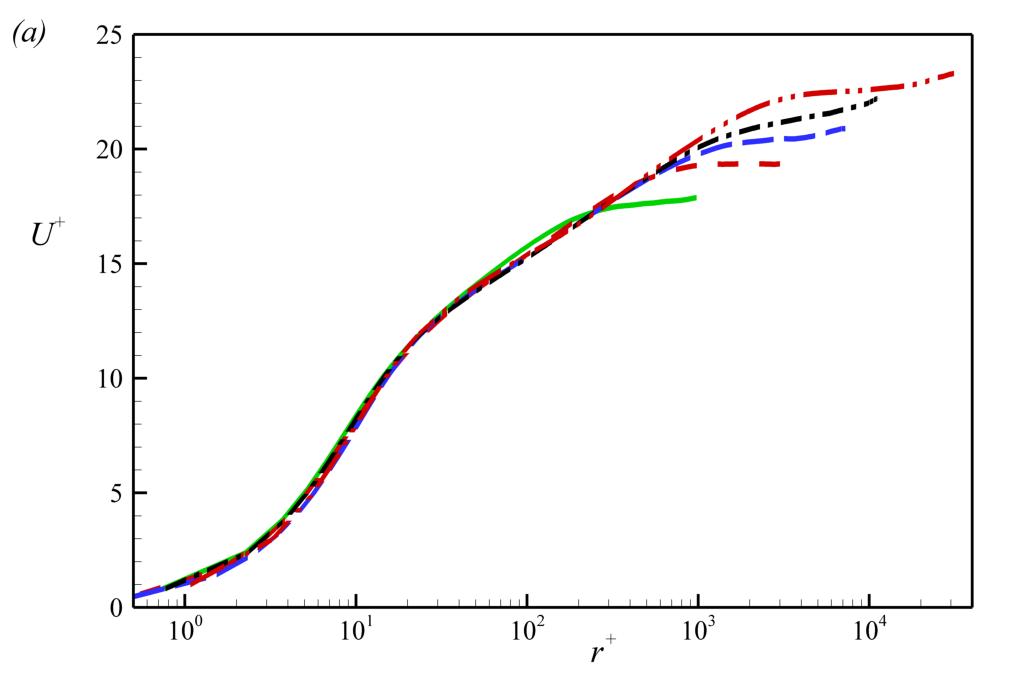}
\includegraphics[width=6.7cm]{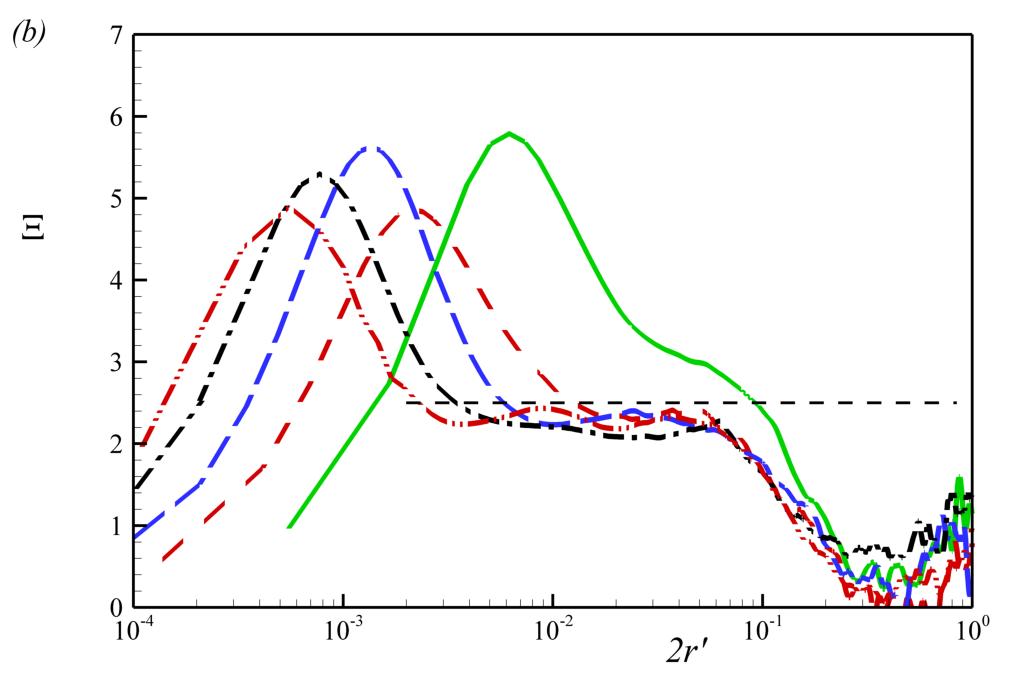}
\caption{ Mean velocity profile for all present LES. (a),  the mean azimuthal velocity versus $r^+$; (b), parameter $\Xi$ versus $2r'$ with $r'=(r-R_i)/d$. Lines for different $Re_i$:
\solid, $ 10^5$,  \dashed;  $3\times 10^5$; \longdashed  $6\times 10^5$;  \chndot, $10^6$; \chndotdot, $3\times 10^6$.}
\label{fig_ut_all}
\end{figure}

As shown for TC flows at $Re_i=10^5$ and $3\times 10^5$, the log-variation in the velocity profile $U^+$ persists only in a range of  $r^+$. When $r^+$ is relatively large, meaning  close to the gap center, the profile deviates substantially from the log law. This can be attributed  to the strong span-wise redistribution effect produced by Taylor vortices which results in an almost constant angular momentum. This will be discussed further subsequently.  In the  estimate of ~\cite{ostilla2016near}, $r^+=0.1 Re_\tau$ is considered as an upper bound for the log layer for $\eta=0.909$.
Mean velocity profiles  $U^+(r^+)$ obtained from LES at higher $Re_i$ are shown in Fig \ref{fig_ut_all}, which plots five lines, representing the five cases implemented.

 Another way to clarify a  possible log region is a scaled parameter which is typically defined as
\begin{equation}
\Xi = r^+ \frac{d U^+}{ dr^+}.
\end{equation}
In the sense of a classic log law, $\Xi$ is equal to the inverse of the K\'arm\'an  constant, $1/\kappa$.
In figure \ref{fig_ut_all}(b), we show plots of $\Xi $ for all  cases. A  horizontal straight line at  $\Xi=2.5$ is also shown, corresponding to $\kappa=0.4$.
It can be observed that $\Xi$ in all higher $Re_i$ LES  extends to about $1\%$ of the half gap, which is consistent with DNS at $Re_i$ up to $3\times 10^5$.


\subsection{Turbulence intensity profiles}

\begin{figure}
\includegraphics[width=6.7cm]{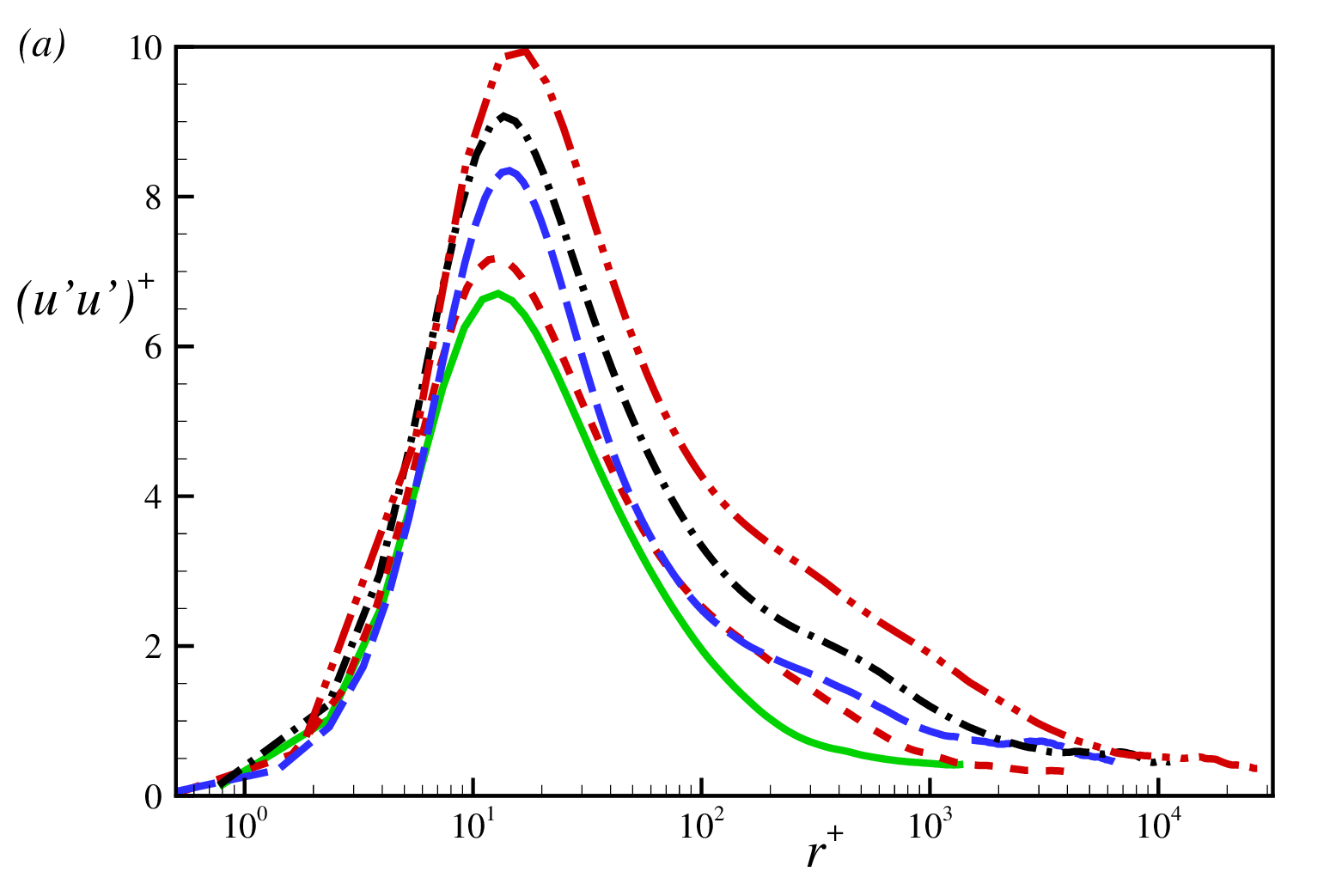}
\includegraphics[width=6.7cm]{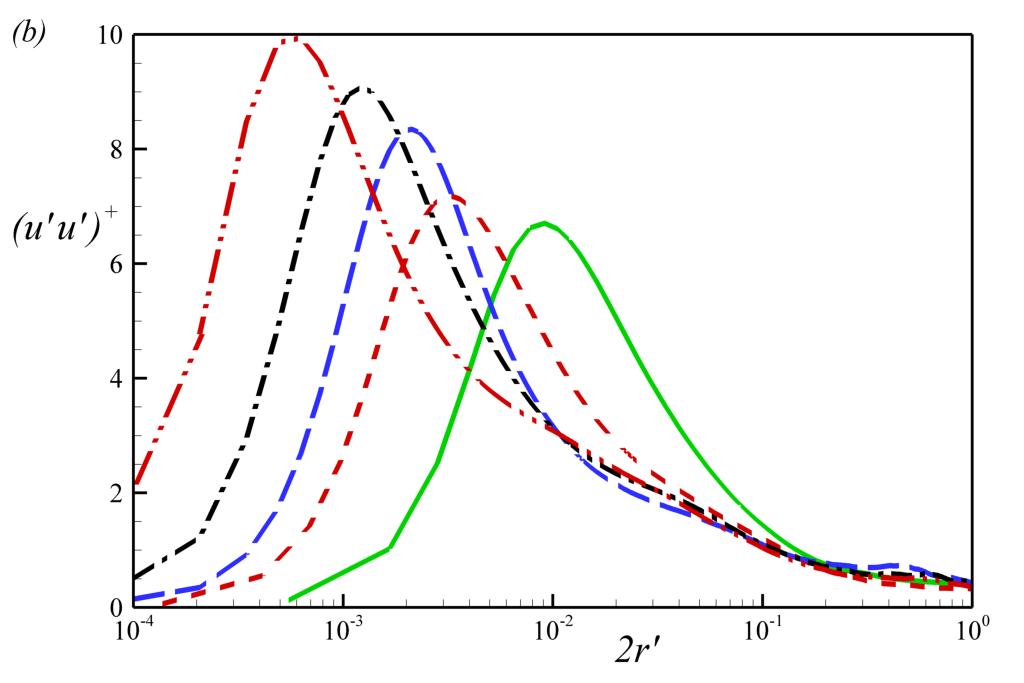}
\caption{ Turbulent intensities for all present LES. (a), versus $r^+$; (b), versus $2r'$.
Lines for different $Re_i$:
\solid, $ 10^5$;  \dashed,  $3\times 10^5$; \longdashed,  $6\times 10^5$;  \chndot, $10^6$; \chndotdot, $3\times 10^6$.
}
\label{fig_fluct_all}
\end{figure}

We consider azimuthal velocity intensities  near the inner cylinder.  In figure \ref{fig_fluct_all}, we show the scaled radial variation of  $(u'_\theta u_\theta')^+$ in two different length scales. Figure \ref{fig_fluct_all}(a)  uses inner scaling in the form $(u'_\theta u_\theta')^+ $ versus $r^+$ ,  while in figure \ref{fig_fluct_all}(b),  the outer scale $r'$ is utilized. For the purpose of readability we  follow usage with channel flow  as $2r'$,  which reaches unity at the gap centerline.  The turbulence intensity profiles display interesting features that are now discussed.

When $(u'_\theta u_\theta')^+ $ is plotted versus $r^+$,  all present LES with $\eta=0.909$ show an inner peak in the range $10<r^+<15$. This  is consistent with the peak location at $r^+\approx 12$ in  TC flow experiments   with $\eta=0.716$ and with Reynolds number up to $1.5\times 10^6$  \cite[]{Huisman2013},
 with a peak location at about  $y^+\approx 15$ in experiments of boundary layer flows up to $Re_\tau=21,430$ \cite[]{Squire2016}  and with super-pipe experiments  up to $Re_\tau=98,187$ \cite[]{Hultmark2012}.
  DNS of channel flow \cite[]{Lee2015} shows a weak increase in the location of the peak value of turbulent intensities
 from $y^+ \approx 15.0$ at $Re_\tau=1,000$ to $y^+ \approx 15.6$ at $Re_\tau\approx5,200$.  In wall units, the present LES does not show a clear tendency for the peak to move outwards  as $Re_{\tau_i}$ increases at our largest values.

In contrast, our LES   does indicate an unambiguous  increasing tendency of the magnitude of the inner-scaled, peak azimuthal intensity,  reaching about $ (u'_\theta u_\theta')^+ \approx    10.0$ at $Re_i =  3\times 10^6$. In experiments of pipe low \cite[]{Hultmark2012}, the peak stream-wise intensity at large  $Re_\tau$ is found to saturate and even decrease, while such saturation is not observed in experiments on zero-pressure-gradient boundary-layer flow \cite[]{hutchins2009}.
 At  similar  $Re_\tau\approx 5,000$, the peak stream-wise intensities   for both channel flow and pipe flow are similar, at around $9.0$,  while the peak in boundary layer flow is  smaller, at about $7.8$; see figure 4(c) of \cite{Lee2015}. For  DNS of plane Couette flow,  \cite{Pirozzoli2014Turbulence} find a peak streamwise intensity $\overline{u'\,u'}=10$ at $Re_\tau\approx 1,000$, and no saturation limit is observed.
 Owing to the presence of Taylor rolls  in both TC and plane-Couette flow, we would expect that these  are more similar to each other than to  canonical pipe/channel flows. The possible saturation in turbulence intensities for TC/PC flows at larger $Re_\tau$ than have been explored to date remains an open question.

 In the numerical study canonical turbulent flow like channel \cite[]{Lee2015} or boundary-layer flow \cite[]{simens2009}, the effect of simulation parameters such as computational domain and mesh size, on  mean velocity and turbulent intensity profiles, has been carefully studied.
  For PC flow and TC flow, it is known that span-wise roll motion can strongly impact the zone of  wall-bounded turbulence. In order to alleviate spurious effects for PC flow,  large computational domains of order $30d$  stream-wise and $8d$  span-wise  are needed \cite[]{Pirozzoli2014Turbulence}, where $d$ is the  flat-plate gap.
\cite{Ostilla2015Effects} investigated the effect of both span-wise and azimuthal domain size on TC flow. They found that
finite-domain effects both on  the structure of the near-wall log region and on  turbulent intensity profiles were generally non-negligible at moderate Reynolds numbers. Again the issue remains to be resolved.

In figure \ref{fig_fluct_all}(b), we plot $(u'_\theta u_\theta')^+$ in  outer-flow scaling.  Profile collapse is observed at around $ r' \approx0.05$.
For  larger $r'$, $(u'_\theta u_\theta')^+$ shows a  plateau, which extends to the centerline $r'=d/2$.
This behavior is different  than that  found in either channel flow or pipe flow.  In the latter, stream-wise turbulent intensities in the near-center region, for example $y/\delta > 0.2$ with $\delta$ the half height in channel flow and the radius in pipe flow,  show  monotonically decreasing behavior in the wall-normal direction away from the wall, approaching a minimum at the centerline.  No obvious plateau region is observed.
 The plateau region of $(u'_\theta u_\theta')^+$  near the centerline region for   TC flow can probably be ascribed to span-wise roll motion which transports and redistributes  angular momentum.

Finally we note that,  owing to   the existence of  span-wise roll motion for the present TC flow, the definition of one-point turbulent intensities is not unambiguous.  We have utilized a diagnostic  that includes only  local small-scale turbulence, and that does not  explicitly recognize the  presence of large-scale roll motion.  According to our present azimuthal intensity metric, a tendency to form a hump and possibly a second peak as the driving Reynolds number  increases up  to $Re_i = 3\times 10^6$,  is not observed. Nor was this found by the DNS of \cite{ostilla2016near} with this same metric up to $Re_i = 3\times 10^5$ where $Re_{\tau_i} = 3,920$.  The
 detailed evaluation and study  of alternative definitions  of turbulent intensity that additionally  includes clarification of the effect of domain size is beyond the scope of the  present study. The issue is interesting as indicated by the substantial variation in  profile shapes indicated in figure 12 of  \cite{ostilla2016near}, for turbulent intensities outside the inner peak in both TC and channel/pipe flow .

\subsection{Angular momentum $L$}

\begin{figure}
\centering
\includegraphics[width=9cm]{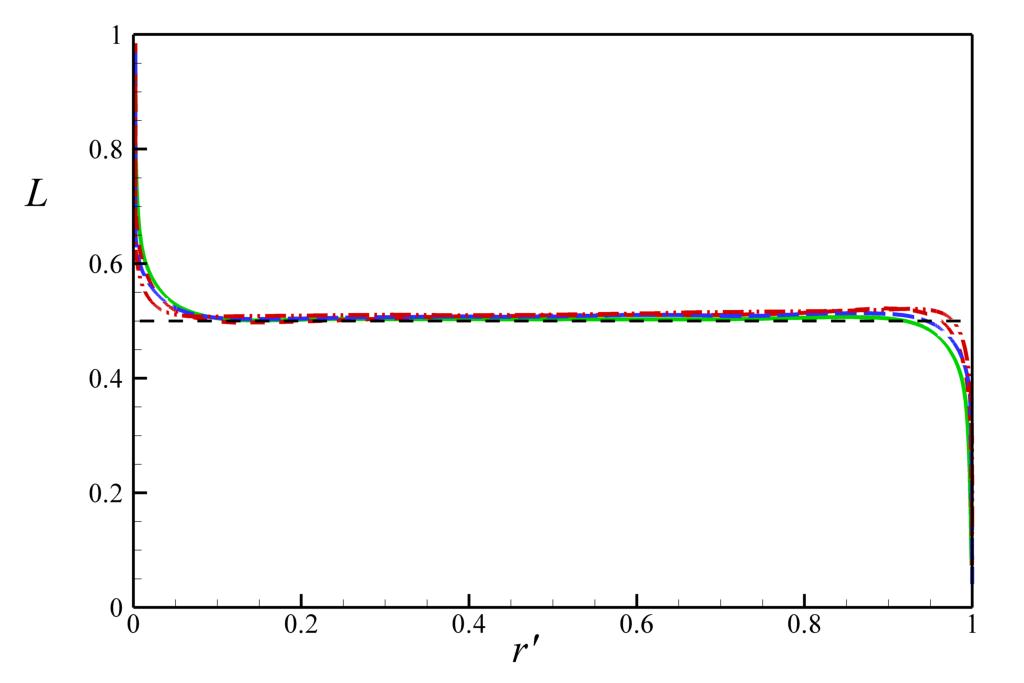}
\caption{Nondimensional angular momentum profiles. $L = r\,U_\theta/(\Omega_i\,R_i^2)$
lines for different $Re_i$: \solid, $10^5$; \dashed, $3\times 10^5$;
 \longdashed, $6\times 10^5$; \chndot, $10^6$; \chndot, $3\times 10^6$.
}
\label{Ang-MomnProfiles}
\end{figure}

The presence of Taylor vortices is thought to transport angular momentum  per unit mass $r\,u_\theta$ between the two cylinders  leading to constant angular momentum in the region separating the two cylinder wall layers ~\cite[]{wereley1999velocity,ostilla2016near}.  The constant is equal to the average of the angular momentum per unit mass of two particles rotating with the angular velocities of both the inner and outer cylinder. With $\Omega_o=0$ this is $L_{av} = \Omega_i\,R_i^2/2$.  Figure  \ref{Ang-MomnProfiles} shows radial profiles of the nondimensional angular momentum $L = r\,U_\theta/(\Omega_i\,R_i^2)$.
A horizontal straight line of $L=1/2$ is also plotted for comparison
For all $Re_i$ shown $L\approx 1/2$ over most of the gap between the two cylinders.

\section{Empirical flow model}
\label{FlowModel}

\subsection{Three-region model}

The constancy of angular momentum across  the cylinder gap with in a region bounded by the two turbulent log-like layers adjacent to the cylinder walls suggests a simple empirical mean-flow model of the present TC flow with the outer cylinder stationary.  The model development to follow is strictly one-dimensional in the radial direction. It predicated on the existence of a finite region of constant angular momentum  $L =   1/2$ for arbitrarily large $Re_i$.
 We first divide the radial domain $R_i\le r\le R_o$  into three regions, denoted I, II, III. In regions I and III the azimuthal mean flow is modeled as wall layers represented by log-like profiles relative to the wall while in the central region, $U$ corresponds to radially constant angular momentum.  The dimensions and mean velocity profiles in the laboratory frame are then  given by

\noindent I:  $R_i\le r \le R_i+\delta_i$ with mean azimuthal velocity:

\begin{equation}
U_\theta = \Omega_i\,R_i - u_{\tau_i}\,\left( \frac{1}{\kappa}\,\ln\left(\frac{(r-R_i)\,u_{\tau_1}}{\nu}\right) +A  \right).
\label{RegionIVel}
\end{equation}

 \noindent  II:  $R_i+\delta_i\le r \le R_o-\delta_o$ with mean azimuthal velocity corresponding to constant angular momentum given by the average of values of the two cylinder angular velocities

 \begin{equation}
U_\theta = \dfrac{1}{2\,r}\,\Omega_i\,R_i^2.
\end{equation}

\noindent  III:  $R_o-\delta_o\le r \le R_o$ with mean azimuthal velocity
\begin{equation}
U_\theta = u_{\tau_o}\,\left( \frac{1}{\kappa}\,\ln\left(\frac{(R_o-r)\,u_{\tau_o}}{\nu}\right) +A  \right).
\label{RegionIIIVel}
\end{equation}
where $u_{\tau_i}$,$\,u_{\tau_o}$ are friction speeds at the inner/outer cylinder surfaces,  $\delta_i$,$\delta_o$ are respectively the thicknesses of the inner and outer cylinder  layers, and $\kappa$ and $A$ are the K\'arm\'an constant and  turbulent boundary-layer offset parameter respectively. The model replaces the wake region, present in pipe and boundary layer flows, by a zone of constant but known azimuthal velocity corresponding to constant angular momentum.

For given $\Omega_i,R_i,R_o,\nu$ there are four unknowns:  $(u_{\tau_i},u_{\tau_o},\delta_i,\delta_o)$.
Two equations can be obtained by matching $U$  at $r = R_i+\delta_i$ and $r= R_i-\delta_o$. A third is the relation, obtained by equality of the magnitude of the  torque exerted at each cylinder surface on the fluid
\begin{equation}
u_{\tau_o} = \eta\, u_{\tau_i}.
\label{TorqueEq}
\end{equation}
The velocity matching equations are
\begin{align}
\Omega_i\,R_i &- u_{\tau_i}\,\left( \frac{1}{\kappa}\,\ln\left(\frac{\delta_i\,u_{\tau_i}}{\nu}\right) +A  \right) - \dfrac{\Omega_i\,R_i^2}{2\,(R_i+\delta_i)} =0,
\label{Vel-Inner}\\
\eta\,u_{\tau_i}\,&\left( \frac{1}{\kappa}\,\ln\left(\frac{\delta_o\,\eta\,u_{\tau_i}}{\nu}\right) +A  \right) - \dfrac{\Omega_i\,R_i^2}{2\,(R_o-\delta_o)} =0.
\label{Eq:OuterVelocity}
\end{align}
where \eqref{TorqueEq} has been used in \eqref{Eq:OuterVelocity}.

For our purposes it will be sufficient to consider \eqref{Vel-Inner}, which is one equation for the two unknowns $(u_{\tau_i},\delta_i)$. A closure relation is required.  The wall layer region is of  thickness $\delta_i$, achieved by the action of the span-wise rolls which mixes the angular momentum to a constant state in region II and perhaps also act to limit the radial growth of the wall layers.  We introduce the assumption that $\delta_i$ scales on $u_{\tau_i}$ and $\Omega_i$ as
\begin{equation}
\delta_i  = K\,\frac{u_{\tau_i}}{\Omega_i},\label{Delta}
\end{equation}
where $K$ is a dimensionless constant that is independent of $\eta$. There are other possibilities, for example scaling the left side of \eqref{Delta} by $d/R_i$.  Equation \eqref{Delta} seems the  most simple and  most physically appropriate.  This can be expressed in a dimensionless form as
\begin{equation}
\frac{\delta_i}{d} = \alpha\, \dfrac{Re_{\tau_i}}{Re_i}\,\dfrac{\eta}{1-\eta},
\label{BLThickness}
\end{equation}
where $\alpha = 2\,K$ and we have used that $R_i = d\,\eta/(1-\eta)$.
Next, we substitute
(\ref{BLThickness}) into  \eqref{Vel-Inner} to obtain a single equation for $Re_{\tau_i}$ when other parameters are specified.  After some algebra, this can be expressed in nondimensional form as
\begin{align}
\kappa( Re_i^2   - 4\,\alpha\,A\,Re_{\tau_i}^2  &+2\,(\alpha - 2\,A)\,Re_i\,Re_{\tau_i} )\nonumber\\
 & -4\, Re_{\tau_i}\,(Re_i+ \alpha\, Re_{\tau_i}) \ln
   \left(\frac{2\,\alpha\,\eta  Re_{\tau_i}^2}{Re_i\,(1-\eta)}\right)=0.
  \label{ReTauEqAlpha}
\end{align}
In  \eqref{ReTauEqAlpha} $\kappa,A$ can be chosen as standard log-law parameters but $\alpha$ is a model-dependent parameter. When these are specified together with $\eta$  and $Re_i$, (\ref{ReTauEqAlpha})  can be solved numerically for  $Re_{\tau_i}$. Then $\delta_i/d$ can be calculated from \eqref{BLThickness}.  Once the parameters of the inner-cylinder wall layer are known, then  \eqref{Eq:OuterVelocity}  can be used to determine the single remaining parameter $\delta_o/d$.

In the sequel we will choose $A=4.5$, $\kappa = 0.4$. In their DNS of TC flow with $\eta = 0.909$, \cite{ostilla2016near}  report $Re_{\tau_i}=1410$ at $Re_i =10^5$. Solving \eqref{ReTauEqAlpha} with these parameters and with $\alpha = 0.25,\, 0.5,\,0.75,\,1.0$ gives $Re_{\tau_i}= 1529,\,1418, \,1360, \,1323$ respectively. For all subsequent calculations with the present model, we will use $\alpha = 0.5$ which gives satisfactory agreement with  DNS for this case.
Setting $\alpha = 1/2$ in   \eqref{ReTauEqAlpha} leads to our basic model equation
\begin{align}
\kappa( Re_i^2 - 2\,A\,Re_{\tau_i}^2  &+  (1 - 4\, A)\, Re_i\, Re_{\tau_i})\nonumber\\
 & -2\, Re_{\tau_i}\,(2\, Re_i+Re_{\tau_i}) \ln
   \left(\frac{\eta  Re_{\tau_i}^2}{Re_i\,(1-\eta)}\right)=0.
   \label{ReTauEq}
\end{align}
According to the structure of the model, the presence of a uniform angular momentum zone separating the two wall layers means that these  behave somewhat independently but are connected by \eqref{TorqueEq}. Numerical calculations show that for smooth walls, to a good approximation $\delta_o = \delta_i/\eta$.

\subsection{Approximate analytical solution}

Equation \eqref{ReTauEq} is not solvable in terms of standard special functions.  A useful   approximation for the $Re_{\tau_i}(\eta,Re_i)$ relation can be obtained by observing that generally $Re_{\tau_i} << Re_i$. This is supported by  experiment, DNS, LES and numerical calculations with  \eqref{ReTauEq}.  Neglecting
the term $2\,A\,Re_{\tau_i}^2$ in the bracketed expression multiplied by $\kappa$ and also the $Re_{\tau_i}$ term in the factor multiplying the log in   \eqref{ReTauEq} and dividing by $Re_i$ then gives
\begin{equation}
\kappa \left(Re_i + Re_{\tau_i}(1 - 4\, A)\right)
 - 4\,Re_{\tau_i} \ln
   \left(\frac{\eta  Re_{\tau_i}^2}{Re_i\,(1-\eta)}\right)=0.\label{ReTauEqSimp}
\end{equation}
This reduction is supported by inspection of the numerical order of magnitude of all  terms in \eqref{ReTauEq} for solutions with parameters in the present range of interest.  This (not shown) indicates that the neglected terms are subdominant. Equation \eqref{ReTauEqSimp} will be seen to  provide a good analytical approximation to exact numerical solutions of \eqref{ReTauEq} over the range of parameters considered presently.

Equation \eqref{ReTauEqSimp} has an analytic solution for $Re_{\tau_i}$ as
\begin{equation}
Re_{\tau_i}(Re_i,\eta) = \frac{\kappa\,Re_i}{8\,W(Z_1)}, \quad\quad
 Z_1 = \frac{ \kappa\,\eta^{1/2}\,Re_i^{1/2}\,\exp[ \kappa(4\,A-1)/8]}{8\,(1-\eta)^{1/2}},
 \label{ReTauEqApprox}
\end{equation}
where $W(Z)$ is the principal branch of the Lambert  (or ProductLog) function, defined as the inverse of $Z = W\,\ln W$.  The Lambert function is sub-logarithmic, with expansion for large $Z$ ~\cite[]{corless1996lambertw}
\begin{align}
W(Z) = L_1&- L_2 +\dfrac{L_2}{L_1} + \dfrac{L_2\,(-2+L_2)}{2\,L_1^2} + \dfrac{L_2\,(-6-9\,L_2+2\,L_2^2)}{6\,L_1^3} \nonumber\\
&+ \dfrac{ L_2\,(-12+36\,L_2-22\,L_2^2+3\,L_2^3) }{12\,L_1^4}+ O\left(\left(\dfrac{L_2}{L_1}\right)^5\right), \nonumber\\
&L_1 = \ln (Z),\quad\quad L_2 = \ln(\ln(Z)).
\label{LambertWFn}
\end{align}

Some results are shown in table \ref{table:CompDNS} in comparison with the DNS of Ostilla-M\'onico \emph{et al.} (2016) and also with  the results of the present wall-resolved LES.  Results using both \eqref{ReTauEq} and \eqref{ReTauEqApprox} are shown, where  differences in  calculated values of $Re_{\tau_i}$ are less than $0.5$\%.
Other values of $\eta$ in the range $0.5-0.91$ show similar errors in the approximate versus exact numerical model estimates of $Re_{\tau_i}(\eta,Re_i)$.
Also shown are calculations for both $\delta_0/d$ and $\delta_i/d$, where $\delta_0=\eta\,\delta_i$.  Equations
\eqref{ReTauEqApprox} and \eqref{BLThickness} show that $\delta_i/d$ decreases slowly with increasing $Re_{\tau_i}$ as the reciprocal of the Lambert function with argument proportional to $Re_{\tau_i}^{1/2}$.

\begin{table}
\begin{center}
\begin{tabular}{p{1.5cm}p{1.5cm}p{1.5cm}p{2.0cm}p{2.0cm}p{1.5cm}p{1.5cm} }
$Re_i$ &  $Re_{\tau_i}$ &$Re_{\tau_i}$  & $Re_{\tau_i}$&$Re_{\tau_i}$  &$\frac{\delta_i}{d}$ &$\frac{\delta_o}{d}$   \\
              &  DNS  & WR-LES   &Eq. \eqref{ReTauEq}  & Eq. \eqref{ReTauEqApprox}   &  &                                                                                                                                                                 \\  \hline
$1\times10^5$ & $1410 $ &     1400          &$ 1418$&  $1426$ &  $0.0708$ &$0.0779$  \\  
$2\times10^5$ & $2660 $ &  $\quad$                & $ 2633$ &   $2646$           &$0.0657$&$0.0723$\\  
$3\times10^5$ & $3920 $ &    $3908$                 & $ 3788$     &    $3807$             &$0.0631$&$0.0694$ \\  
$6\times10^5$ & $\quad$ &    $7289$                 & $ 7078 $     &    $7112$             & $0.0589$     & $0.0648$   \\  
$1\times10^6$ & $\quad$ & $11250$&    $11246$                 &  $11298$                &$0.0562$& $0.0618$\\  
$3\times10^6$ & $\quad$ & $31780$ &$30622$&    $30751$            &$0.0510$&$0.0561$ \\  
$1\times10^7$ & $\quad$ & $\quad$ &$92583$&    $92916$            &$0.0462$&$0.0508$ \\  
$1\times10^8$ & $\quad$ & $\quad$ &$7.829\times 10^5$&    $7.856\times 10^5$  &$0.0391$&$0.0430$ \\  
$1\times10^9$ & $\quad$ & $\quad$ &$6.765\times 10^6$&    $6.785\times 10^6$  &$0.0338$&$0.0372$ \\  
\end{tabular}
\end{center}
\caption{Model calculations compared with DNS ~\cite[]{ostilla2016near} and the present wall-resolved LES.  Two versions of the model calculations for $Re_{\tau_i}$  are shown in columns 4 and 5. }
\label{table:CompDNS}
\end{table}%


The above model is expected to be valid for $ 0.6\le \eta <1$ but not $(1-\eta)<<1$.  This is because, when $\eta\to1$ with $\Omega_o=0$, the turbulent flow is expected to similar to plane-Couette flow where $\delta_i/d = \delta_o/d \approx 1/2$.  Substitution of   \eqref{ReTauEqApprox} into \eqref{BLThickness} gives $\delta_i/d$ as a function of $(\eta,Re_i)$. When $\eta \to1 $ at fixed $Re_i$ it is found that $\delta_i/d$ diverges, which is nonphysical.  At any fixed $Re_i$ we can calculate the value  of $\eta$ for which $\delta_i/d =1/2$, which may be taken as defining rough limits on the validity of the model.  For $Re_i=10^5,10^6,10^7,10^8,10^{10}$ these values are respectively $\eta = 0.9887,0.9909,0.9923,0.9934,0.9949$.  These are sufficiently close to $\eta =1$ to give confidence that the model is useful for practical TC cylinder radii ratios.

\subsection{Nusselt number approximation}

The Taylor number is defined by equation \eqref{TaNumII}.
The Nusselt number  $Nu$ is defined as the ratio of the torque required to sustain a statistical steady state of turbulent motion  to the torque required for strictly laminar viscous motion at the same $Re_i$~\cite[]{grossmann2016high}.  Using the viscous flow solution and the definition of $u_{\tau_i}^2 = \tau_{w,i}/\rho$, $Nu$ can be expressed, for $\Omega_o=0$ as
\begin{equation}
  Nu = \frac{2\,\eta\,(1+\eta)\,Re_{\tau_i}^2}{Re_i}.
\label{TaNum2  }
\end{equation}
 Hence, for given $\eta$ and $Re_i$,  if $Re_{\tau_i}$ is known from a solution to  \eqref{ReTauEq},  then both $Ta$ and $N_u$ can be calculated.
\begin{figure}[ht!]
\begin{center}
\includegraphics[scale=0.45]{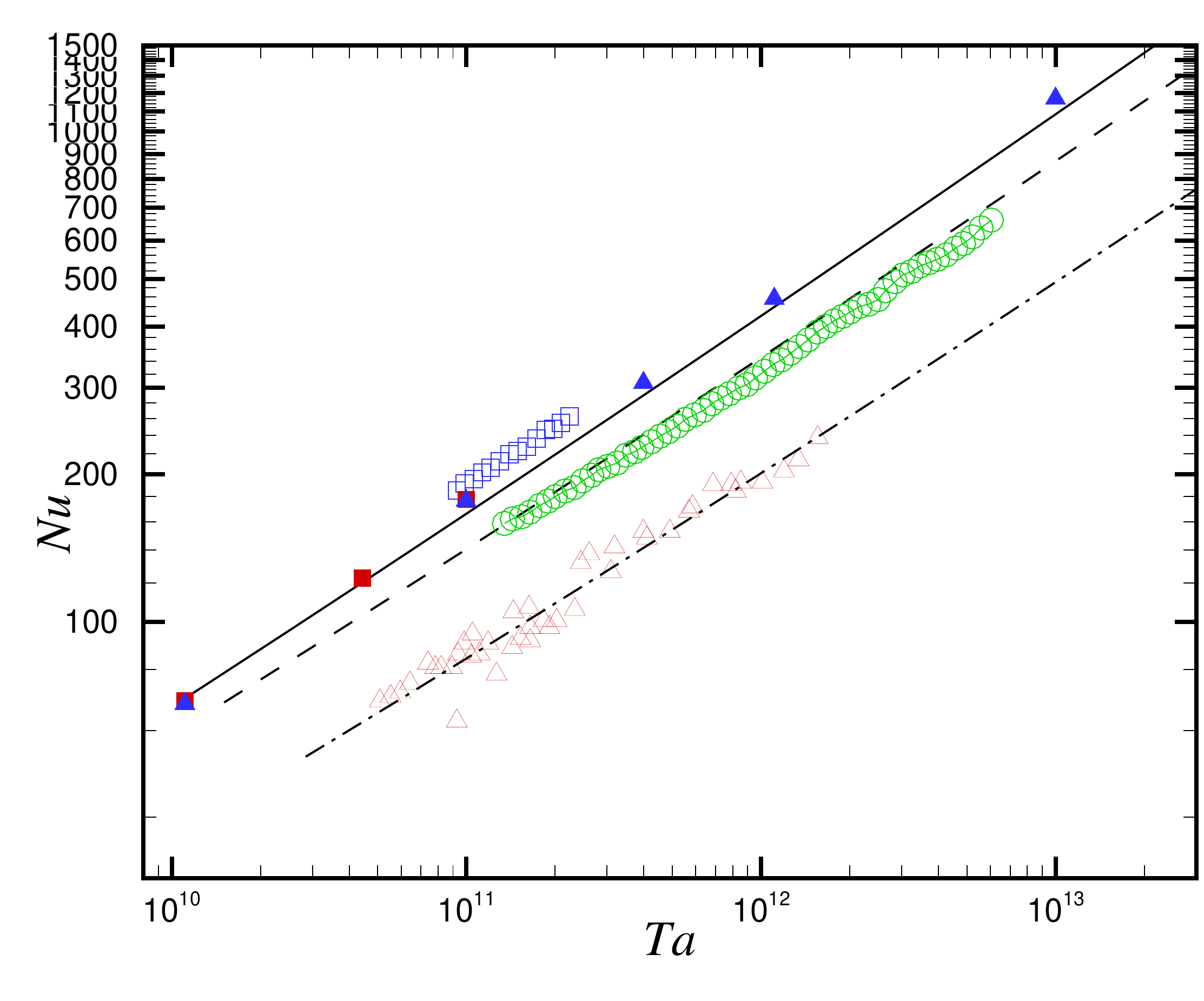}
\caption{\small{  { $Nu$ versus $Ta$.
Open symbols; experiment $\eta = 0.909$ (\square),
$0.72$($\circ$) ~\cite{van2011torque,van2012optimal}; $\eta=0.5$ ($\triangle$)~\cite{merbold2013torque}.
 $\sq$; DNS of $\eta = 0.909$ by~\cite{ostilla2016near} .
 $\blacktriangle$; present LES of $\eta = 0.909$.
 Lines:  from \eqref{Nu-TaEqApprox}.  \solid; $\eta=0.909$.
 \dashed; $\eta = 0.72$.
 \chndot;$\eta = 0.5$.
} }}
\label{Fig1}
\end{center}
\begin{center}
\includegraphics[scale=0.45]{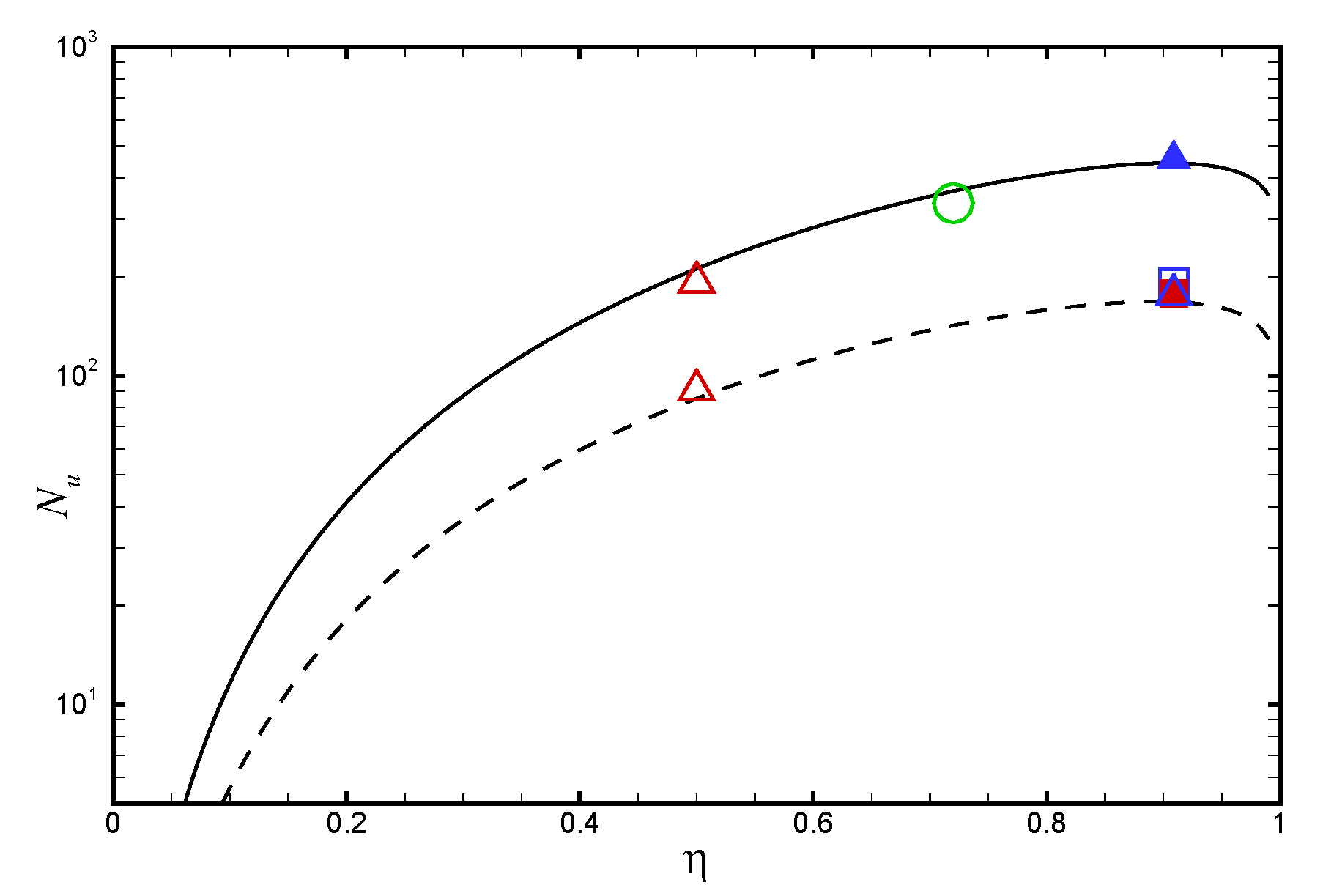}
\caption{\small{$Nu$ versus $\eta$. Dashed line $Ta= 10^{11}$. Solid line $Ta= 10^{12}$. Symbols key; see figure \ref{Fig1} }}
\label{Fig1A}
\end{center}
\end{figure}
  Alternatively, when  \eqref{ReTauEqApprox} is combined with \eqref{TaNum2  }, this gives
\begin{equation}
Nu(Ta,\eta) = \frac{\kappa^2\,\eta^3\,Ta^{1/2}}{4\,(1+\eta)^2\,(W(Z_2))^2}, \quad\quad
 Z_2 = \frac{ \kappa\eta^{3/2}\,Ta^{1/4}\,\exp[ \kappa\,(4\,A-1)/8]}{2^{3/2}(1-\eta)^{1/2}\,(1+\eta)^{3/2}}.
 \label{Nu-TaEqApprox}
\end{equation}
Specific calculations show, that for the range of $\eta$ and $Ta$  considered presently, numerical solutions of  \eqref{ReTauEq} together with \eqref{TaNum2  } agree with  \eqref{Nu-TaEqApprox} to $1$\% or better, improving with increasing $Ta$.

  Figure \ref{Fig1} shows $Nu$ verses $Ta$ for $\eta$ fixed  using \eqref{Nu-TaEqApprox} compared with DNS ~\cite[]{ostilla2016near}, our wall-resolved LES for $\eta = 0.909$ and with data for
   $\eta = 0.5,0.72,0.909$~\cite[]{merbold2013torque,van2011torque,van2012optimal} obtained from  ~\cite{grossmann2016high}, while figure  \ref{Fig1A} shows $Nu$ versus $\eta$ for two values of $Ta= 10^{11}, 10^{12}$.  The model appears to capture well the dependence of $Nu(Ta,\eta)$ on both $Ta$ and $\eta$ over the range shown. The decrease of $Nu$ with $\eta$ larger than about $\eta = 0.91$ may not be physically correct and may indicate the limitation of the model when $\eta\to1$. It is clear from the analytic form and the known behavior of the Lambert function, that $Nu(\eta,Ta)$ with $\eta$ fixed increases more slowly than $Ta^{1/2}$.  Using \eqref{LambertWFn},  \eqref{Nu-TaEqApprox}  has the leading order asymptotic form
\begin{equation}
Nu(Ta,\eta) = \frac{\kappa^2\,\eta^3\,Ta^{1/2}}{4\,(1+\eta)^2\,(\ln(Z_3)-\ln(\ln(Z_3)))^2} + HOT, \quad\quad
 \label{Nu-TaEqApproxAsyA}
\end{equation}
For gigantic $Ta$, this becomes
\begin{equation}
Nu(Ta,\eta) = \frac{4\,\kappa^2\,\eta^3\,Ta^{1/2}}{(1+\eta)^2\,(\ln[Ta])^2}  + HOT. \quad\quad
\label{Nu-TaEqApproxAsyAA}
\end{equation}
Equation  \eqref{Nu-TaEqApproxAsyAA} is not a good approximation to \eqref{Nu-TaEqApprox}  at $Ta$ typical of the highest $Ta$ experimental data. Power law behavior for $Nu(\eta,Ta)$ has been proposed (see  ~\cite{grossmann2016high}). A power-law approximation to \eqref{Nu-TaEqApprox} may be a good fit over a few decades in $Ta$ but, according to the present model, this cannot represent  the correct very large $Ta$ asymptote.

\subsection{Angular momentum profiles}

It is straightforward to calculate profiles of the angular  momentum $L$  from the model. When normalized such that $ L  = rU_\theta /(\Omega_i\,R_i^2)$, this gives, with  $r' = (R_i-r)/d$

\noindent I:  \;\; $
L = \left(\frac{(1-\eta ) r' }{\eta }+1\right) \left(1-\frac{2 \,Re_{\tau_1}}{Re_i}\,
   \left( \frac{1}{\kappa}\,\ln [2 \,Re_{\tau_1} \,r']+ A \right)\right),\quad 0\le r \le\delta_i /d,  $

\noindent II: \;\;  $  L = 0.5,\quad  \delta_i /d \le r' \le 1- \delta_0/d,$

\noindent  III:\;\;  $
L = \left(\frac{(1-\eta ) r'}{\eta }+1\right) \,\frac{2 \,\eta\,Re_{\tau_1}}{Re_i}\,
   \left( \frac{1}{\kappa}\,\ln [2\,\eta \,Re_{\tau_1} \,(1- r')]+ A \right),\;\;  1-\delta_o/d \le r'  \le1.  $

\begin{figure}
\centering
\includegraphics[width=6.5cm]{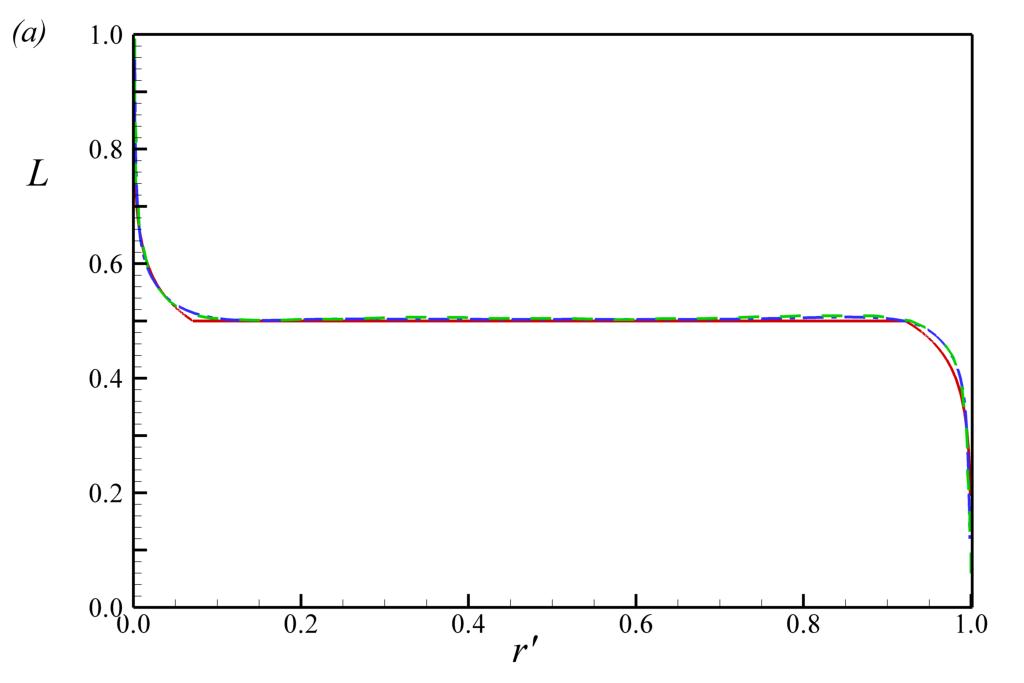}
\includegraphics[width=6.5cm]{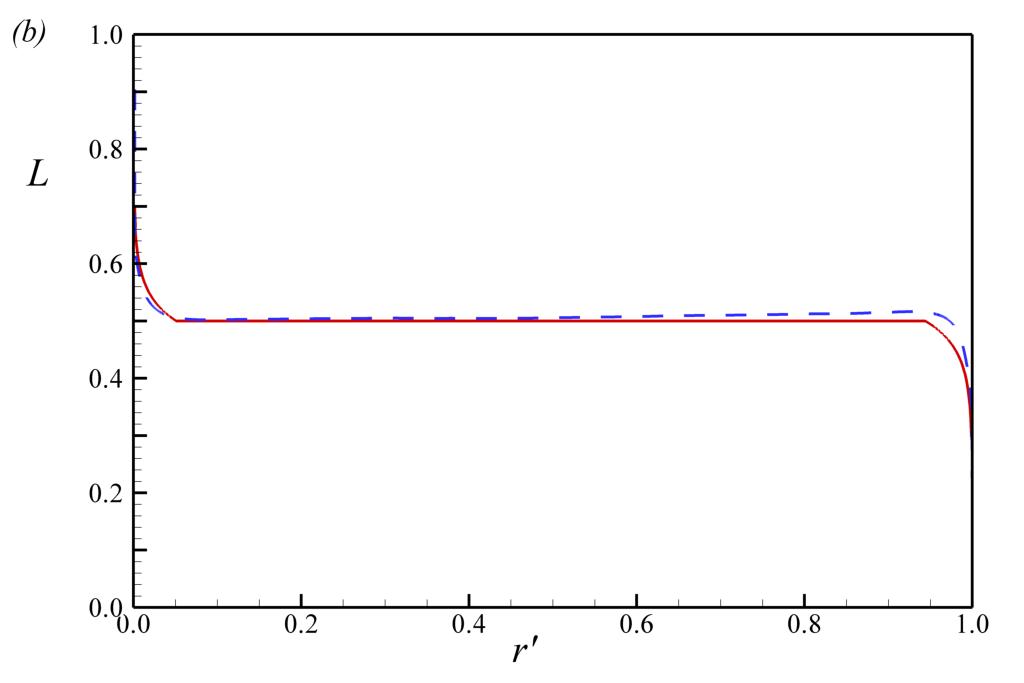}
\caption{Radial angular momentum profiles. Model compared with DNS and LES. (a); $Re_i = 10^5$.
 (b), $Re_i = 3\times10^6$. \solid, model prediction; \dashed, present LES;
   \chndot, DNS by \cite{ostilla2016near}.}
\label{AngMomnComparison}
\end{figure}

Radial angular momentum profiles calculated from these expression are compared with DNS and the present LES in
figure \ref{AngMomnComparison}. The agreement is satisfactory.

\subsection{Boundary layer thickness}

\begin{figure}
\centering
\includegraphics[width=6.5cm]{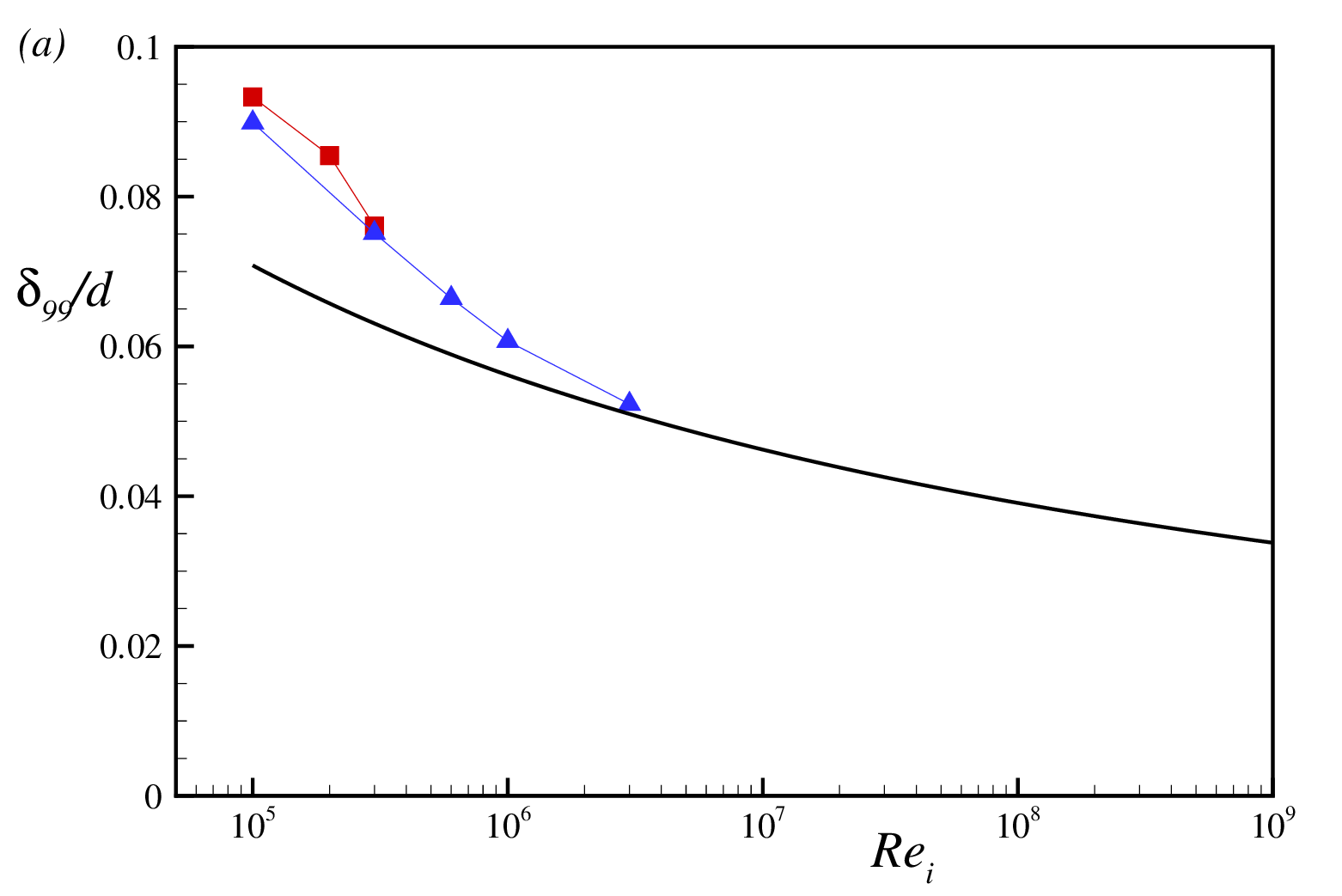}
\includegraphics[width=6.5cm]{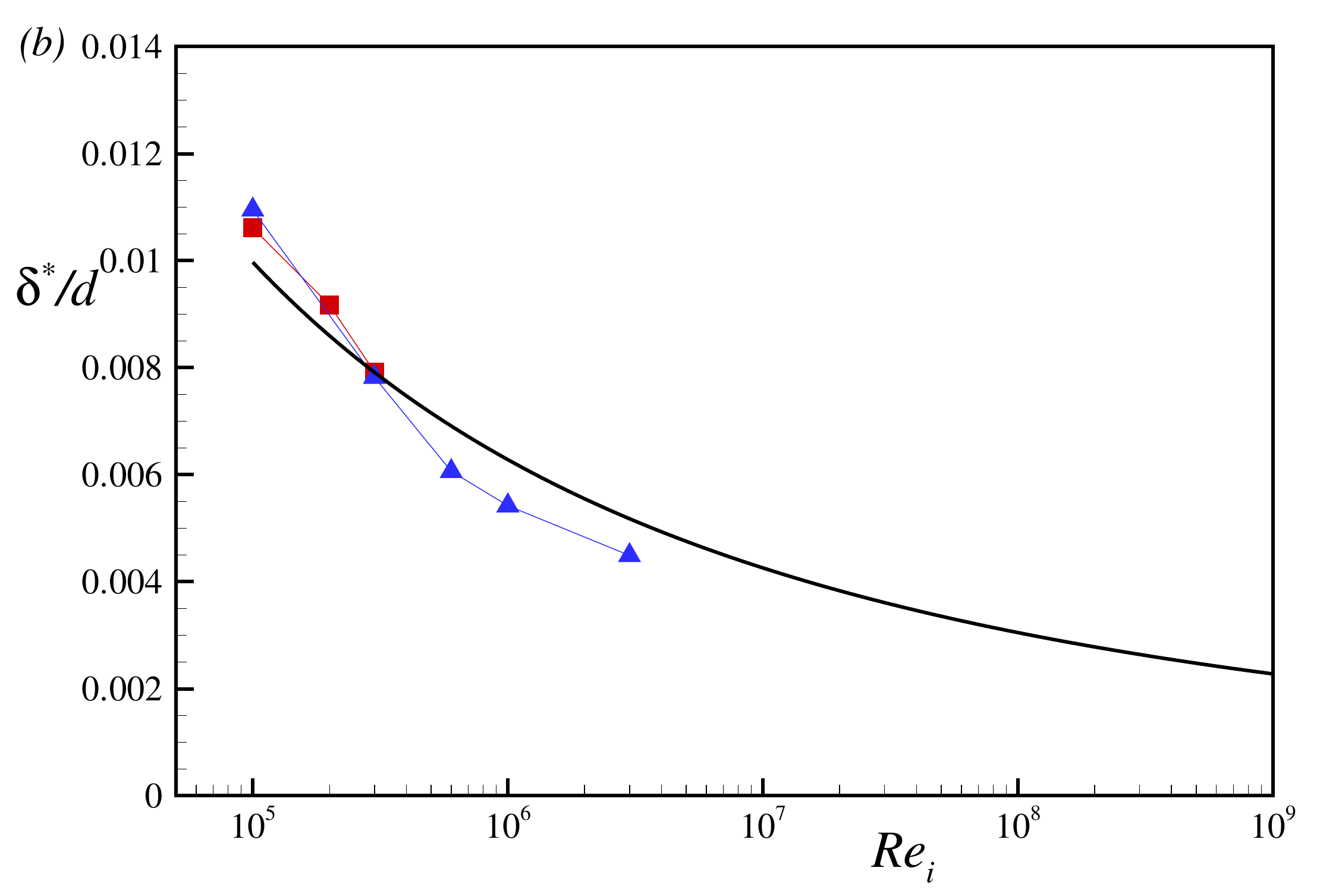}
\includegraphics[width=6.5cm]{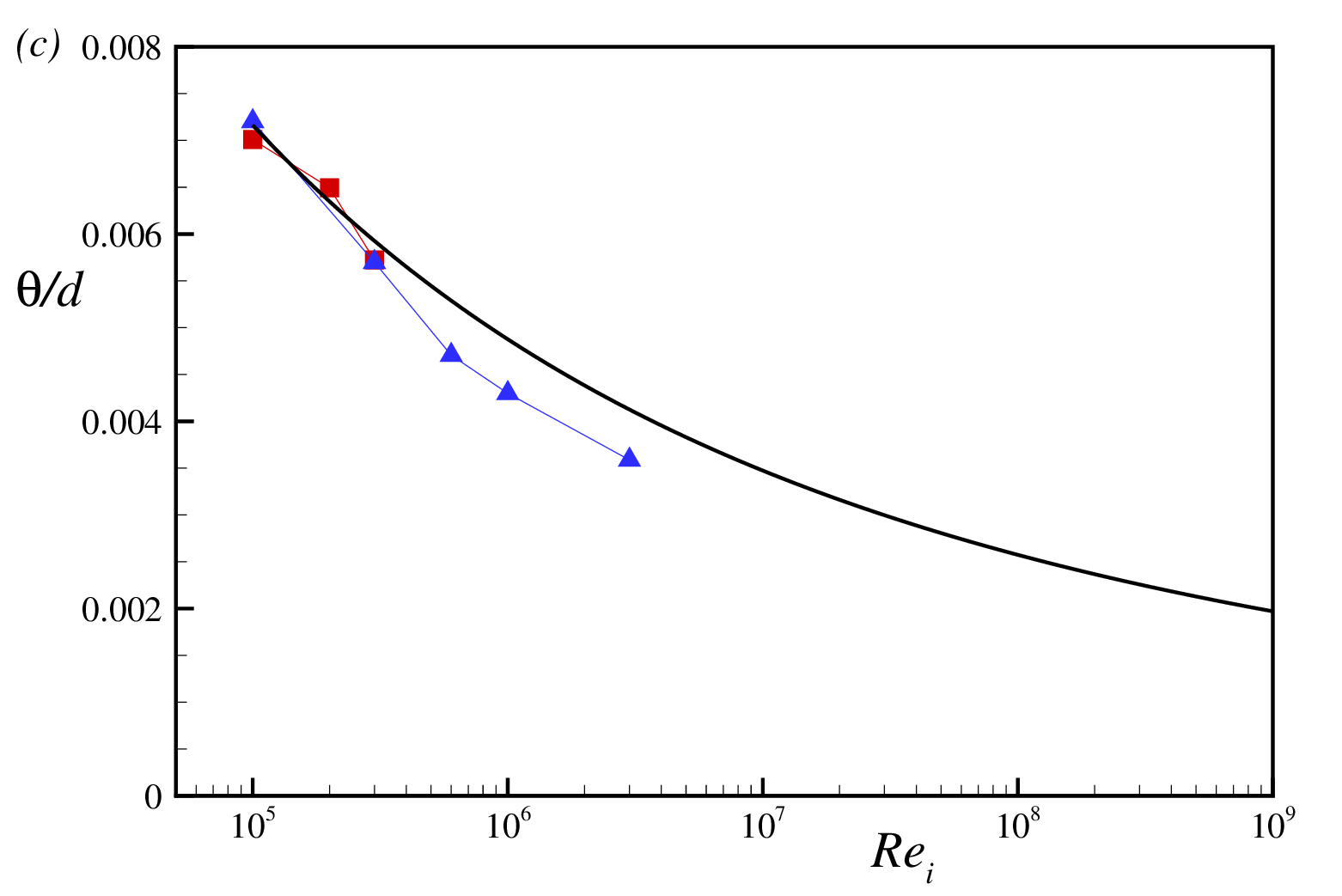}
\includegraphics[width=6.5cm]{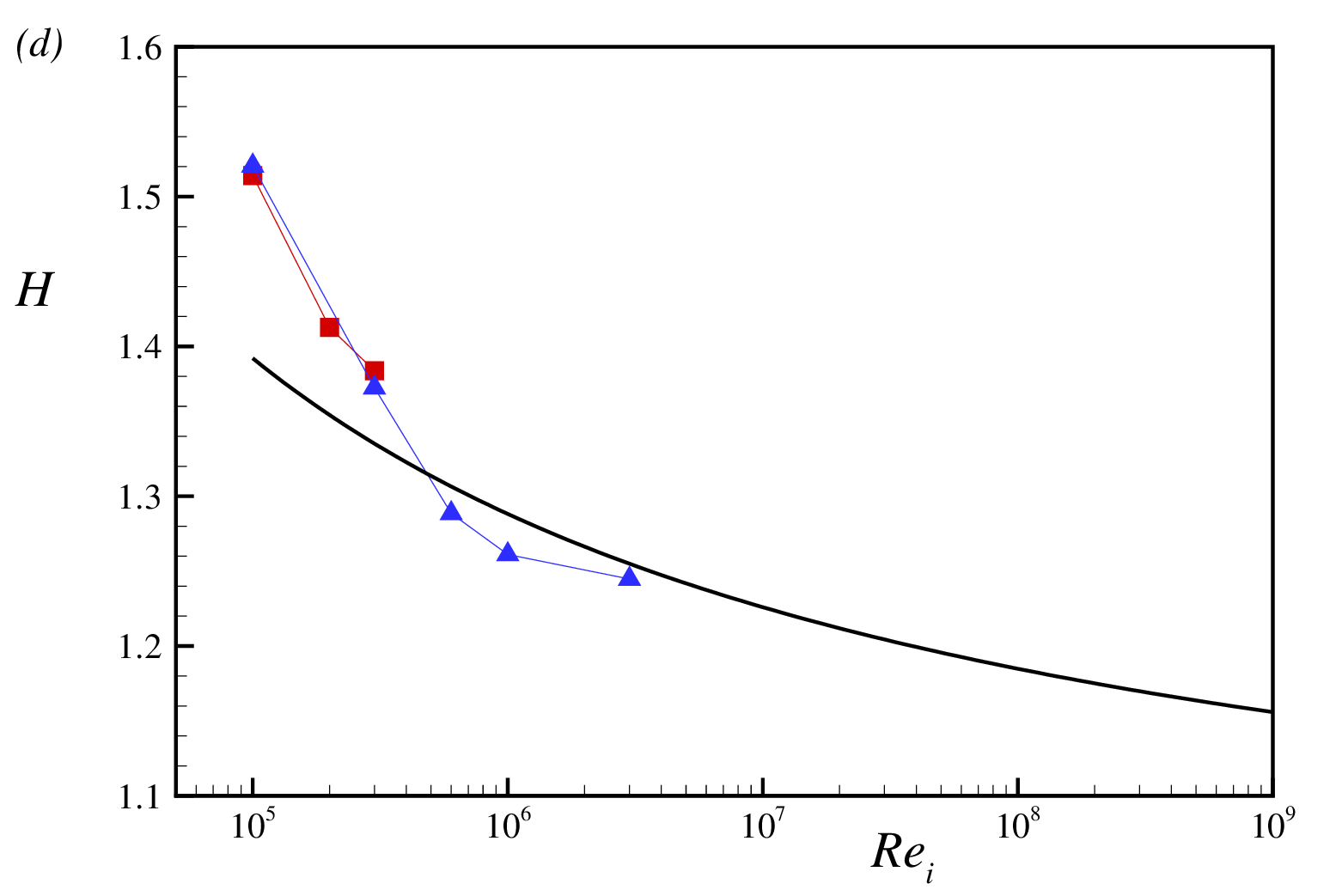}
\caption{Measures of Boundary layer thickness and shape factor $H$ parameter versus $Re_i$.
(a), $\delta_{99}/d$; (b), $\delta^*/d$; (c), $\theta/d$; (d), $H$.
Solid line; present model. Symbols: $\sq$, DNS by~\cite[]{ostilla2016near}; $\blacktriangle$ LES. }
\label{BLWidth}
\end{figure}

Once the parameters $Re_{\tau_i}$ and $\eta_i$ have been determined, the velocity profile in the log-regions I and II can be calculated.  This allows calculations of the displacement and momentum thicknesses as functions of $Re_{\tau_i}$.  For the inner cylinder, these  are defined presently as
\begin{equation}
\delta^* = \int_0^{\delta_i}\left(1 - \frac{U(r)}{U_L} \right)\,dr,\quad\quad \theta = \int_0^{\delta_i}\frac{U}{U_L}\left(1 - \frac{U(r)}{U_L} \right)\,dr
\end{equation}
where $U_L =  \Omega_i\,R_i- \textstyle{\frac{1}{2}}\,\Omega_i\,R_i^2/(R_i+\delta_i)$.  Using the log-part of the velocity profile in \eqref{RegionIVel} , these expressions can be evaluated analytically.  The resulting expressions are cumbersome and details are omitted presently.  Both  $\delta^*$ and $\theta$ can be calculated from DNS and LES.  An issue is the upper cutoff in the integrations. Presently this was determined as  $\delta_{99} =R_i-r$ where $r$  satisfies
\begin{equation}
\dfrac{ U(r) - (\Omega_i\,R_i - 0.5\,\Omega_i\,R_i^2/r)}{(\Omega_i\,R_i - 0.5\,\Omega_i\,R_i^2/r)} <0.01.
\end{equation}
Results for $\delta_{99},\delta^*,\theta$ and the shape factor $H= \delta^*/\theta$ from the model are shown in figure \ref{BLWidth} in comparison with both DNS and LES.  Identifying $\delta_i$ with the measured $\delta_{99}$ provides an over estimate. Both DNS and LES indicate a decline in the respective measures of wall-layer thicknesses as $re_i$ increases, in agreement with the model.

\subsection{Discussion}

It is of interest to discuss the state of flow for arbitrarily large $Re_i/Ta$. Because the DNS/LES appear to be in good agreement with the model for mean-flow properties, we will consider the large $Re_i$ limit of the model behavior.  It is clear that the model indicates that  when $Re_i/Ta$  increase at  fixed $\eta$ for smooth-wall flow on both cylinder surfaces,  $\delta_i/d$ and $\delta_o/d$  decline as the inverse of the Lambert $W$-function. This follows from \eqref{ReTauEqApprox} and \eqref{BLThickness}. If the constant $L$ region separating the two wall layers remains intact at exceptionally large $Re_i$ or $Ta$, this suggests a limiting mean flow consisting of two asymptotically thinning wall layers in relation to the cylinder gap $d$, separated by a region where
\begin{equation}
U_\theta = \dfrac{1}{2\,r} \,\Omega_i\,R_i^2.
\label{UthetaLimit}
\end{equation}

The present model does not contain a description of turbulent intensities. But both DNS and LES appear to support the hypothesis that $ {
{u'_\theta\,u'_\theta}}$ and other intensities scale  on $u_{\tau_i}^2$. Since $u_{\tau_i}/(\Omega_i\,R_i) \sim Re_{\tau_i}/Re_i$ must also decline (this follows from \eqref{ReTauEqApprox}) as $Re_i$ increases, then it follows that $ {  {u'_\theta\,u'_\theta} }/(\Omega_i\,R_i)^2$ must also decrease slowly. Hence the present results may be interpreted to imply that the very large $Re_i$ state consists of a mean flow over most of the cylinder gap described by \eqref{UthetaLimit} together with slowly declining turbulent intensities in relation to the square of the driving cylinder surface speed $\Omega_i\,R_i$.  Huge  $Re_i$ would be required to access this asymptotic state.

\section{Rough walls}
\label{RoughWalls}

 It is of general interest to develop the extension of the present empirical mode to turbulent  rough-wall flow with sand-grain-type roughness of scale $k_s$. We represent the effects of sand-grain roughness by use of the mean-velocity offset as represented by the Colebrook roughness function
\begin{equation}
\Delta^+ U(k_s^+) = \dfrac{1}{\kappa}\,\ln\left( 1 + \beta k_s^+  \right), \quad\quad \beta = \exp\left( \kappa\,(A-B) \right)
\label{Colebrook}
\end{equation}
where $B=8.5$ is a standard constant and  $k_s^+=k_s\,u_{\tau_i}/\nu$.   The functional form for $\beta$ in \eqref{Colebrook} guarantees that when $k_s$ is very large, the expression for the velocity profile is asymptotic to the standard fully rough profile form $u = u_\tau\,(\log((r-Ri)/k_s)/\kappa +B)$~\cite[]{Jimenez2004}. The Colebrook  $\Delta U^+(k_s)$ by no means represents all roughness types but can be taken as typical of the transition from fully smooth, $k_s^+\to 0$,  to fully rough, $k_s^+ >100$,  near-wall behavior.  Again we can treat the inner wall in isolation provided  that the uniform angular momentum region exists, separating the wall layers on the inner and outer cylinder walls.  For the inner wall, the velocity matching equation \eqref{Vel-Inner} is replaced by
\begin{equation}
\Omega_i\,R_i - u_{\tau_i}\,\left( \frac{1}{\kappa}\,\ln\left(\dfrac{\delta_i\,u_{\tau_i}}{\nu} \right) +A -  \frac{1}{\kappa}\,\ln\left(1+ \beta\, k_s^+ \right)\right) - \dfrac{\Omega_i\,R_i^2}{2\,(R_i+\delta_i)} =0. \label{Vel-InnerRough}
\end{equation}
 Equation \eqref{BLThickness} is retained. Again, a single equation for $Re_{\tau_i}$ can be obtained by substituting \eqref{BLThickness} into  \eqref{Vel-InnerRough} with $\alpha=1/2$ and converting to non-dimensional parameters
\begin{equation}
-2\, A\, Re_{\tau_i} + \dfrac{Re_i\, (Re_i + Re_{\tau_i} )}{2 Re_i + Re_{\tau_i}}
- \dfrac{2\,Re_{\tau_i}}{\kappa}\, \ln\left(\dfrac{\eta\, Re_{\tau_i}^2}{(1-\eta)\,Re_i\,  (1+ 2\,\beta\,\epsilon\,Re_{\tau_i})}\right) =0
\label{EqRoughness}
\end{equation}
where $\epsilon = k_s/d$ is the ratio of the sand-grain roughness scale to the cylinder gap.  When $\beta = 0$,  \eqref{ReTauEq} is recovered.

To illustrate the behavior with rough walls, it is preferable to utilize the skin-friction coefficient defined as
$C_f \equiv 2\,\tau_{i,w}/(\Omega_i\,R_i)^2$ where $\tau_{i,w} = \rho\,u_{\tau_i}^2$.    In terms of other parameters $C_f$ can be expressed as
\begin{equation}
C_f = 8\,\dfrac{Re_{\tau_i}^2}{Re_i^2} = \dfrac{4\,Nu}{\eta\,(1+\eta)\,Re_i}.
\label{CfRelations}
\end{equation}
Substituting an expression for $Re_{\tau_i}$ obtained from the   first of \eqref{CfRelations} into    \eqref{EqRoughness}  gives, after some algebra
\begin{align}
2 &- \dfrac{8}{8 + \sqrt{2}\,\sqrt{C_f}}\nonumber \\
&-\dfrac{\sqrt{2\,C_f}}{\kappa}\,
   \left(A\,\kappa + \ln\left(\dfrac{\eta\,C_f\,Re_i}{4\,(1-\eta)\,(2+\sqrt{2\,C_f}\,\epsilon\,Re_i\,\exp (\kappa\,(A-B)))}\right)\right)=0.
   \label{CfEquation}
\end{align}
Two limits are of interest.  The first is the smooth-wall case  $\epsilon\to 0$  with $Re_i$ fixed. This follows directly by putting $\epsilon=0$ in \eqref{CfEquation}, The second is  $Re_i\to \infty$ at any finite $\epsilon>0$, which takes the form
\begin{equation}
2 - \dfrac{8}{8 + \sqrt{2}\,\sqrt{C_f}} -
  \dfrac{\sqrt{2\,C_f}}{\kappa}\,
   \left( B\,\kappa + \ln\left(\dfrac{\eta\,\sqrt{C_f}}{4\,\sqrt{2}\,\,\epsilon\,(1-\eta)}\right)\right)=0.
   \label{CfEquationLim}
\end{equation}
 Hence for fully rough-wall, turbulent wall layer flow, the skin friction, and therefore the torque required to sustain the motion becomes independent of $Re_i$, and depends only on $\eta$ and $\epsilon$.

Neither  \eqref{CfEquation} nor \eqref{CfEquationLim} can be solved analytically for $C_f$. But if the first two terms of the left-hand side  of  \eqref{CfEquationLim}  are replaced by their leading-order Taylor expansion in the small quantity $\sqrt{C_f}$, we obtain
\begin{equation}
1 + \dfrac{\sqrt{C_f}}{4\,\sqrt{2}} - \dfrac{\sqrt{2\,C_f}}{\kappa}\,
   \left( B\,\kappa + \ln\left(\dfrac{\eta\,\sqrt{C_f}}{4\,\sqrt{2}\,\,\epsilon\,(1-\eta)}\right)\right)=0.
   \label{CfEquationAPP}
\end{equation}
This equation has the  solution
\begin{equation}
C_f  = \dfrac{\kappa^2}{2\,W^2(Z)},\quad\quad Z = \dfrac{\kappa\,\eta\,\exp(B\,\kappa-\kappa/8)}{8\,\epsilon\,(1-\eta)}.
\label{CFSoln}
\end{equation}
For the range of parameters considered presently, numerical solutions to  \eqref{CfEquation} agree with \eqref{CFSoln}  to 3--4 significant figures.
\begin{figure}
\begin{center}
\includegraphics[width=6.5cm]{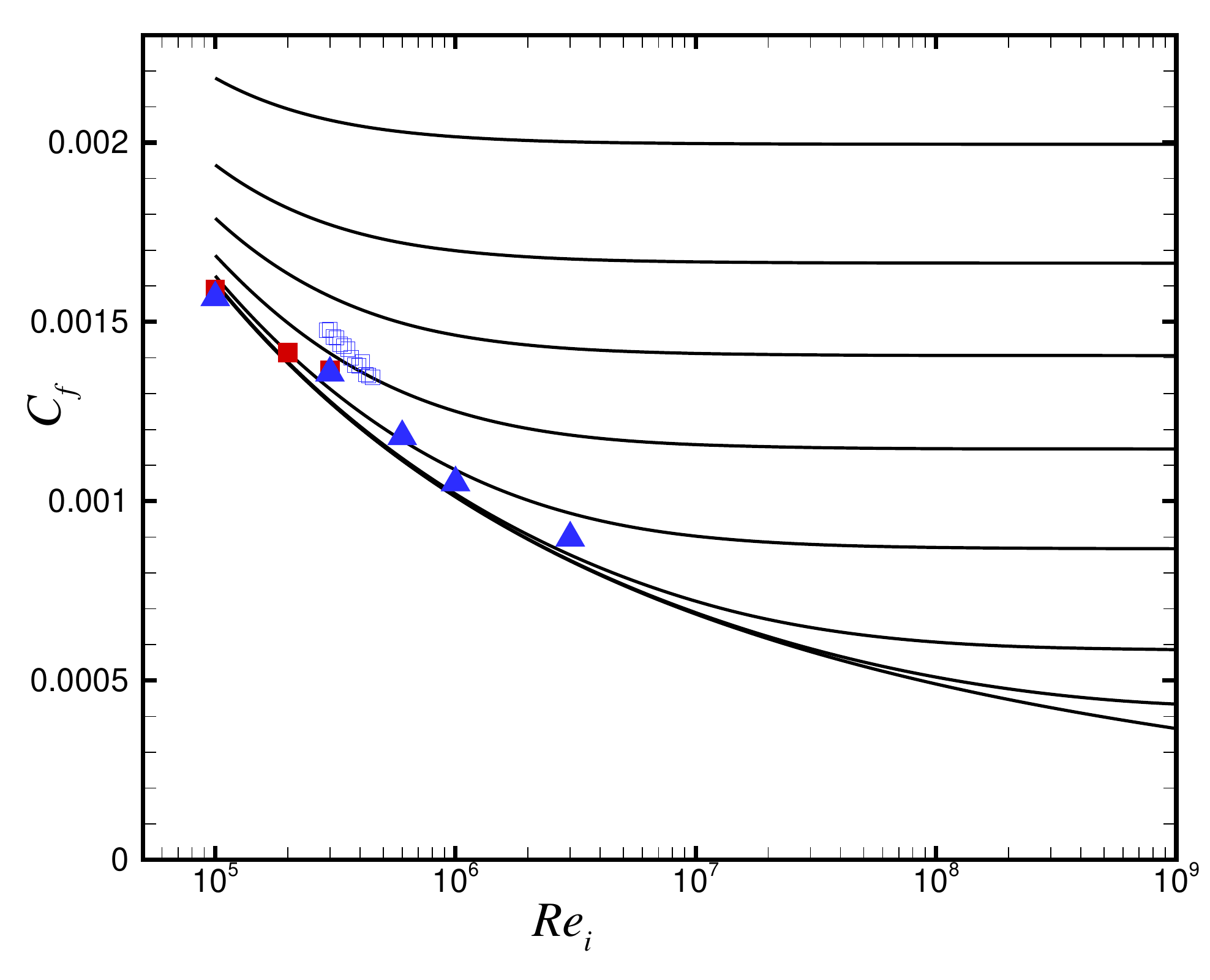}
\includegraphics[width=6.5cm]{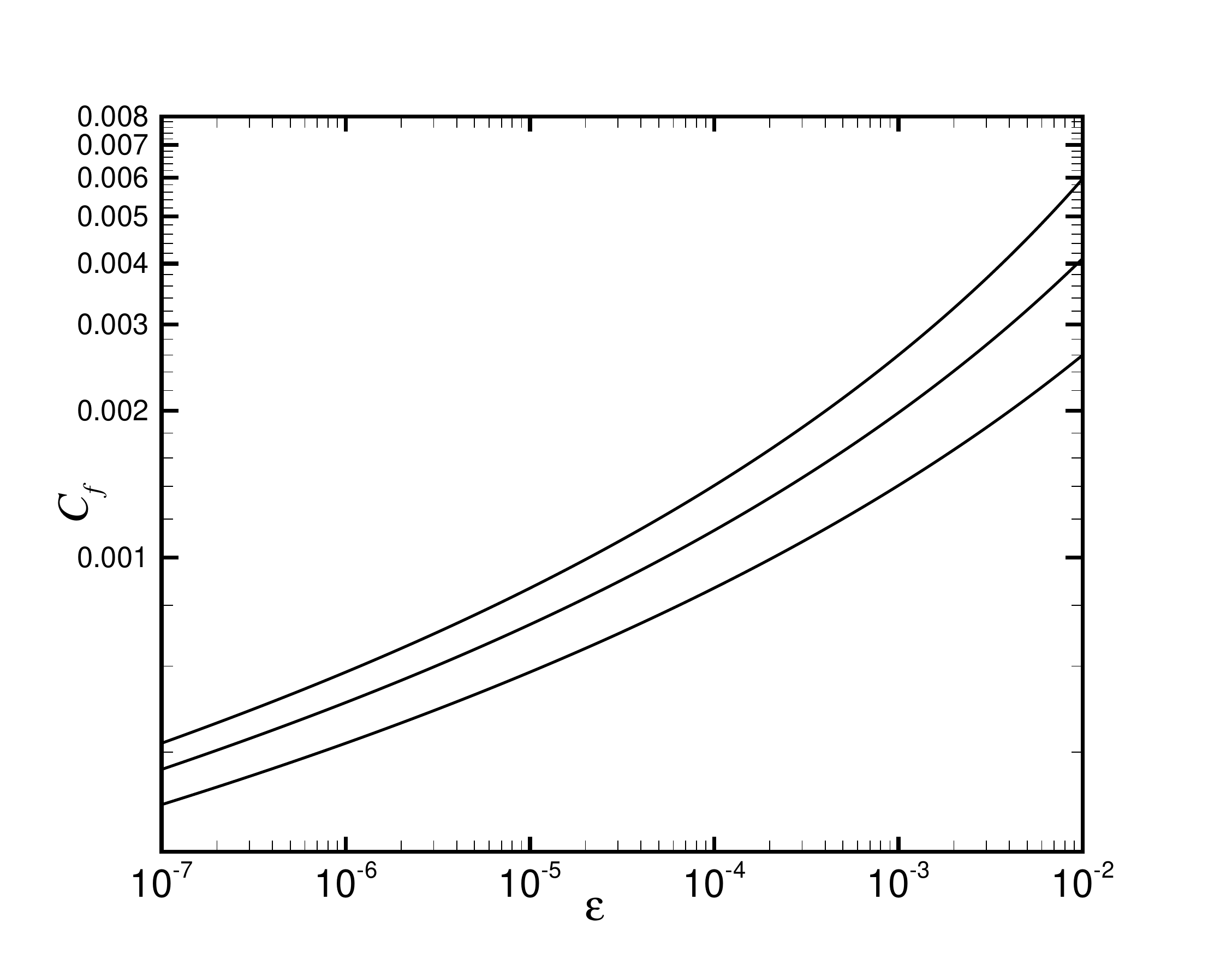}
\caption{\small{ Skin-friction coefficient $C_f$ for different roughness level.
Left :  $C_f$ versus $Re_i$ for $\eta=0.909$.  Solid lines: model prediction. Top to bottom $\epsilon = k_s/d = 4\times 10^{-3},2\times 10^{-3}, 10^{-3},4\times 10^{-4},10^{-4}, 10^{-5},10^{-6},0$.
$\sq$, DNS by \cite{ostilla2016near};  $\blacktriangle$, present LES.  Right:  $C_f$ versus $\epsilon = k_s/d$.  Top to bottom $\eta = 0.5,0.72,0.909$
}}
\label{SkinFriction}
\end{center}
\hspace{0.4in}
\begin{center}
\includegraphics[width=6.5cm]{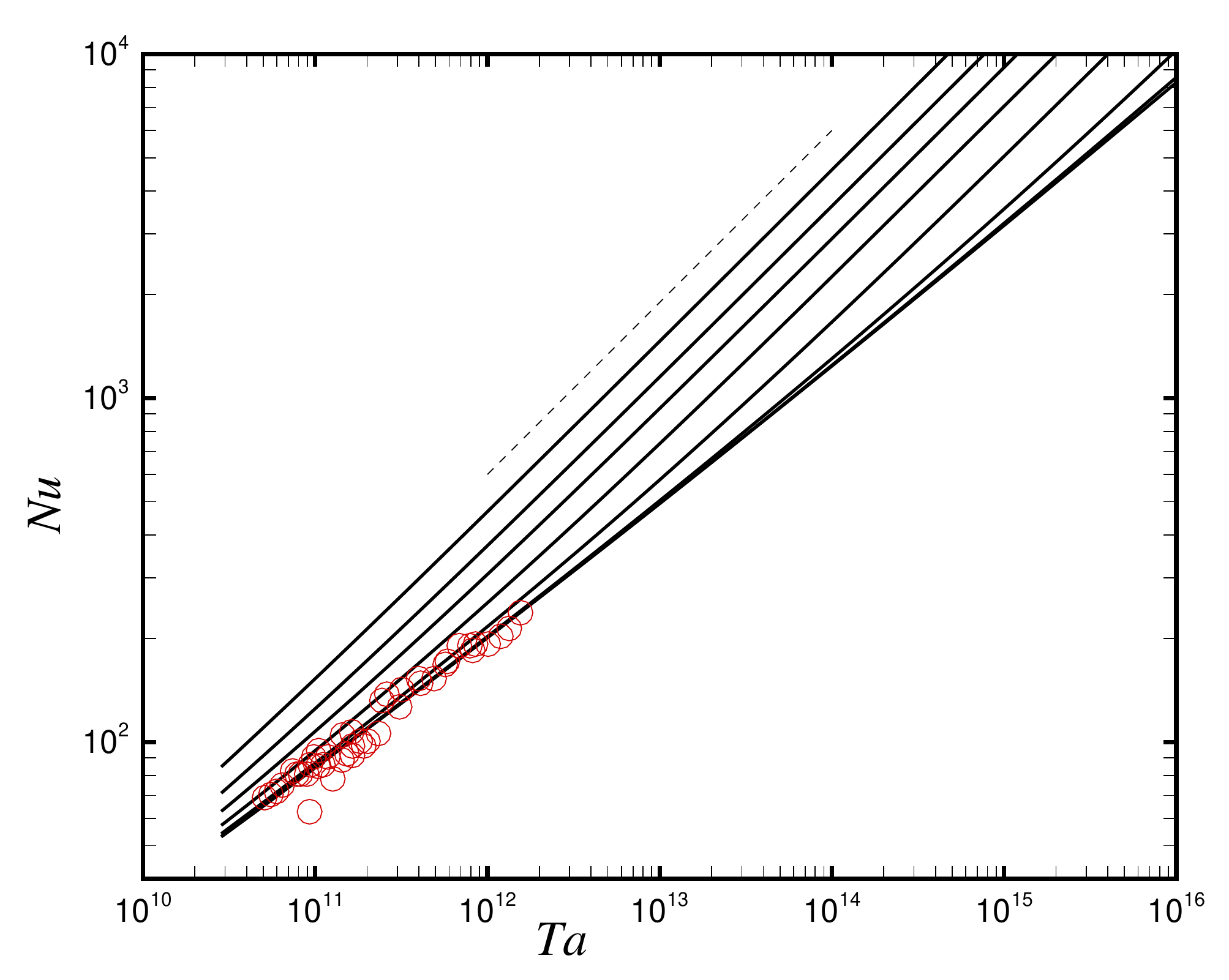}
\includegraphics[width=6.5cm]{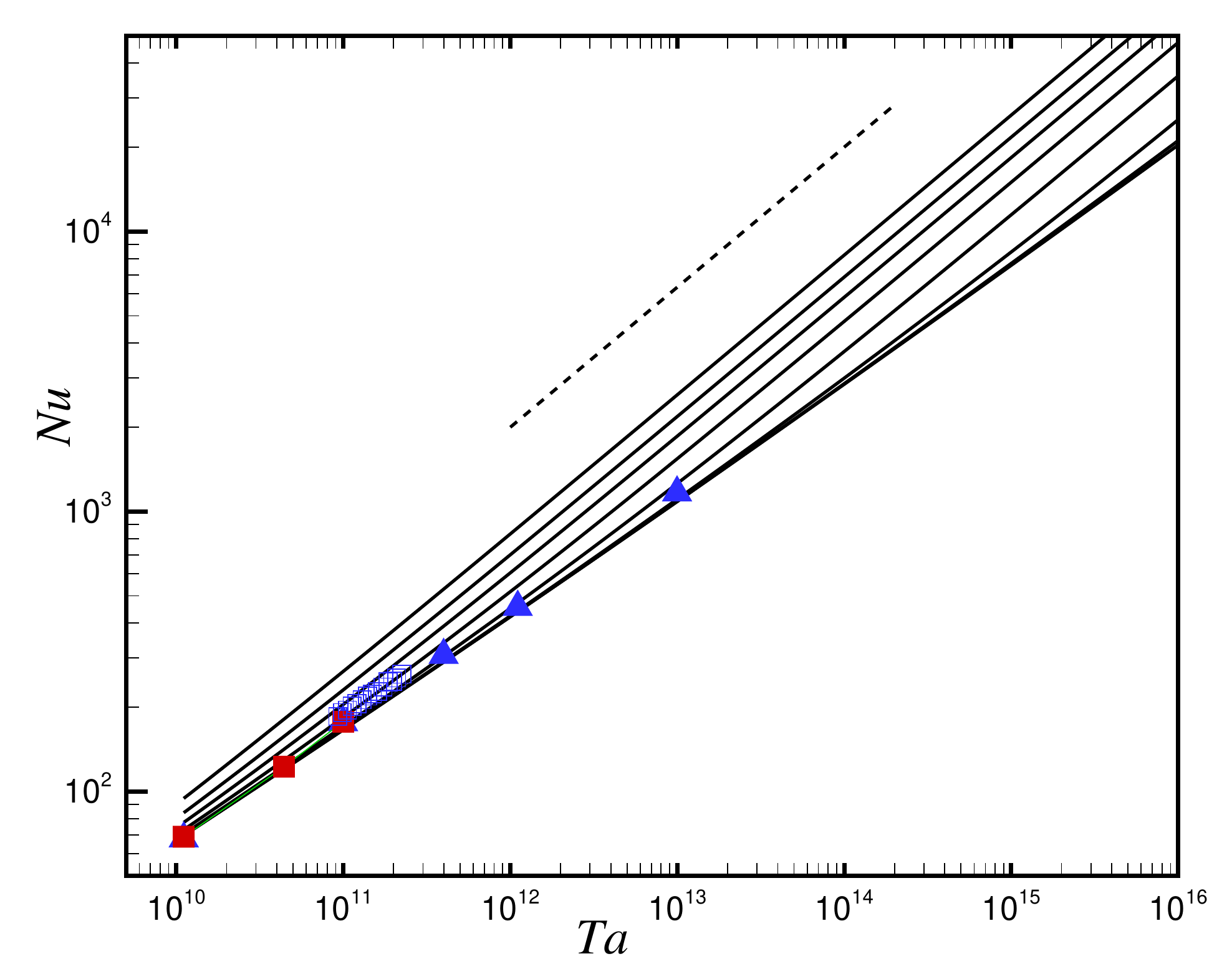}
\caption{\small{$Nu$ versus $Ta$ for rough, inner cylinder walls. Left $\eta=0.5$, right $\eta = 0.909$.
 Solid lines: model prediction.
Dashed line; slope $1/2$. Symbol key; see figures \ref{Fig1} and \ref{SkinFriction}.
}}
\label{NuTaRough}
\end{center}
\end{figure}
Other relevant quantities can now be calculated as
\begin{equation}
 k_s^+ = \dfrac{1}{\sqrt{2}}\,\epsilon\,\sqrt{C_f}\,Re_i,\quad\quad \dfrac{\delta_i}{d} = \dfrac{\eta\,\sqrt{C_f}}{4\,\sqrt{2}\,(1-\eta)}.
\label{ksdeltaieq}
\end{equation}
Together with  \eqref{CfEquationAPP}, the  second of \eqref{ksdeltaieq} shows that when $Re_i\to\infty$, $\delta_i/d$ is also independent of $Re_i$.  The model predicts that, at sufficiently high $Re_i$  and sufficiently small $\epsilon$, the asymptotic rough-wall state consists of constant $C_f$ and wall-layer thickness  $\delta_i/d$ that are independent of $Re_i$ (and hence of $Ta$).  Figure  \ref{SkinFriction} (left) shows $C_f$ versus $Re_i$ with $\eta = 0.909$ with several values of $\epsilon$ including the smooth-wall limit $\epsilon\to 0$, obtained from numerical solution of \eqref{CfEquation}.  This is essentially a Moody diagram for a TC flow with a uniformly rough inner cylinder and an outer stationary cylinder.  The right panel of figure { \ref{SkinFriction}}  shows the fully-rough $C_f(\eta,\epsilon)$ given by \eqref{CFSoln}.

 The large $Re_i$ limit behavior at finite $\epsilon$ is clear. With $\eta = 0.909$,  the second of \eqref{ksdeltaieq} shows that  $\delta_i/d = 1.766\,\sqrt{C_f}$. Hence figure  \ref{SkinFriction} with the ordinate rescaled  also shows the large $Re_i$ behavior of $\delta_i/d \sim \sqrt{C_f}$. For $\epsilon= 10^{-3},10^{-4}, 10^{-5},10^{-6}$, the limiting values are
 $\delta_i/d = 6.62 \times 10^{-2}, 5.20\times 10^{-2}, 4.27\times 10^{-2} , 3.62\times 10^{-2}$ respectively.
   We expect the model to be physically reasonable provided that $\delta_i/k_s>10$ approximately so that
   the log-like wall layer can exist. This is satisfied by all numerical solutions presented.

 For completeness we also show rough-wall model results in $Nu,Ta$ variables  Using \eqref{CfRelations},  $Nu$ is proportional to $Re_i$ with a coefficient proportional to $C_f$, and that depends on both $\eta$ and $\epsilon$.  Using \eqref{TaNumII} and \eqref{CfRelations} we can obtain generally
 \begin{equation}
 Nu = \dfrac{2\,\eta^3\,C_f}{(1+\eta)^2}\,Ta^{1/2}
 \end{equation}
Hence in the fully rough limit it  follows that $Nu \sim Ta^{1/2}$, again with a coefficient that depends on both $\eta$ and $\epsilon$. Figure \ref{NuTaRough} shows model results for $Nu$ versus $Ta$ for rough walls. The transition from smooth wall flow where $Nu \sim Ta^{1/2}$ with Lambert-function corrections to fully rough behavior $Nu \sim Ta^{1/2}$ is clear.

\section{ Conclusion }
\label{Conclusion}

The present study uses wall-resolved large-eddy simulation (LES) to simulate
Taylor-Couette flow with a narrow gap (radius ratio  $\eta= r_i/r_o=0.909$)
between the inner, rotating cylinder and the outer stationary cylinder.
The LES  implemented via a general curvilinear coordinate code with a fully staggered velocity mesh. Fourth-order central difference schemes are used for all spatial discretization.

 Two cases
at $Re_i=10^5$ and $3\times 10^5$ are used as verification cases. By comparing
mean velocity profile $U^+$ and turbulent intensities $(u'_\theta u'_\theta)^+$, $(u'_y u'_y)^+$ and $(u'_r u'_r)^+$, we show that the present LES framework can reasonably
capture the salient features of TC flows, including the quantitative behavior of span-wise Taylor rolls, the log profile in the mean velocity profile and the angular momentum redistribution due to the presence of Taylor rolls, up to $Re_i=  3\times10^6$, which corresponds to a Taylor number $Ta= 9.969\times 10^{12}$.

A simple empirical model is developed for the mean-flow properties of Taylor-Couette flow when the outer cylinder is stationary. The model consists of three contiguous flow regions; two contain turbulent wall layers, one at each cylinder wall while the third is a central, annular region of constant angular momentum. The model requires that this constant angular momentum per unit mass is known and equal to one half of that corresponding to rotation with the inner cylinder angular velocity $\Omega_i$. It is supposed that this three-region state is produced by redistribution of angular momentum by either mean-flow or  instantaneous, fluctuating Taylor-roll motion, and further, that this persists to arbitrarily high Taylor number. Inside each wall layer, the flow is modeled by a standard log-like profile with $\kappa=0.4, A=4.5$. The model takes an analytic form by implementing equality of azimuthal velocity at the region boundaries. It is closed by an additional assumption that the inner boundary-layer thickness is proportional to the local friction velocity divided  by the cylinder angular velocity. This introduces a single arbitrary parameter which is set equal to $0.5$.

 The composite model is shown to capture  the effects of both  the cylinder ratio $\eta$ and  the Taylor number $Ta$ over the range of available DNS, experiment and the present LES.  At large $Ta$ an approximate but sufficiently accurate model reduction  gives a specific analytical form where the Nusselt number grows somewhat slower than the square root of the Taylor number.  This growth is not of power-law form. As $Ta$ increases  both wall layers shrink in thickness. An asymptotic state is indicated  where the uniform angular momentum region occupies almost all of the cylinder gap, with asymptotically small turbulence intensities.

 The model is extended to a rough inner wall comprising uniform sand-grain roughness. Use of a Colebrook-type roughness function allows construction of a Moody-diagram for Taylor-Couette flow. For given $\eta$ and ratio of sand-grain roughness to cylinder gap, an asymptotic rough-wall state is found  with constant  skin friction and boundary-layer thickness that is independent of $Re_i-Ta$.  Here the Nusselt number is proportional to $Ta^{1/2}$.

\smallskip

\section*{Acknowledgement}
 This work was partially supported by the KAUST baseline research funds of R.S..
The Cray XC40, Shaheen,  at KAUST was utilized for all the reported LES.

\bibliographystyle{jfm}
\bibliography{TC-LES}

\end{document}